\documentclass{nature}

\newcounter{firstbib}

\usepackage{graphicx}
\usepackage{epsfig}
\usepackage[font={small}]{caption}
\usepackage{hyperref}
\usepackage{amsmath}
\usepackage{amssymb}
\usepackage{lineno}
\usepackage{tikz}
\tikzset{every picture/.style={/utils/exec={\sffamily}}}

\usepackage[symbol]{footmisc}

\usepackage{xcolor}

\def\gtorder{\mathrel{\raise.3ex\hbox{$>$}\mkern-14mu
             \lower0.6ex\hbox{$\sim$}}}
\def\ltorder{\mathrel{\raise.3ex\hbox{$<$}\mkern-14mu
             \lower0.6ex\hbox{$\sim$}}}

\def\Angstrom{\textup{\AA}}

\def\farcs{\hbox{$.\!\!^{\prime\prime}$}}

\newcommand{\aap}{Astron. Astrophys.}
\newcommand{\araa}{Ann. Rev. Astron. Astrophys.}
\newcommand{\apj}{Astrophys. J.}
\newcommand{\aj}{Astron. J.}
\newcommand{\apjl}{Astrophys. J. Lett.}
\newcommand{\apjs}{Astrophys. J. Suppl.}
\newcommand{\nat}{Nature}

\newcommand{\pasp}{Publ. Astron. Soc. Pac.}

\newcommand{\mnras}{Mon. Not. R. Astron. Soc.}

\newcommand{\prl}{Phys. Rev. Lett.}

\newcommand{\ssr}{Space Sci. Rev.} 

\title{The pair-instability origin of supernova 2023vbw}

\author{
Daichi~Hiramatsu$^{\ref{af:UF},\ref{af:CfA},\ref{af:IAIFI}*}$,
Edo~Berger$^{\ref{af:CfA},\ref{af:IAIFI}}$,
Daichi~Tsuna$^{\ref{af:CfA},\ref{af:TAPIR},\ref{af:RESCEU}}$,
Sebastian~Gomez$^{\ref{af:UT},\ref{af:CfA}}$,
Harsh~Kumar$^{\ref{af:CfA},\ref{af:IAIFI}}$,\\
Peter~K.~Blanchard$^{\ref{af:CfA},\ref{af:IAIFI}}$,
Walter~W.~Golay$^{\ref{af:CfA}}$,
Anya~E.~Nugent$^{\ref{af:CfA}}$,
Takashi~J.~Moriya$^{\ref{af:NAOJ},\ref{af:SOKENDAI},\ref{af:Monash}}$,
D.~Andrew~Howell$^{\ref{af:LCO},\ref{af:UCSB}}$,
Alexei~V.~Filippenko$^{\ref{af:UCB}}$,
Thomas~G.~Brink$^{\ref{af:UCB}}$,
WeiKang~Zheng$^{\ref{af:UCB}}$,
Yi~Yang$^{\ref{af:Tsinghua}}$,
Moira~Andrews$^{\ref{af:LCO},\ref{af:UCSB}}$,
K.~Azalee~Bostroem$^{\ref{af:UA},\ref{af:Catalyst}}$,
Joseph~Farah$^{\ref{af:LCO},\ref{af:UCSB}}$,
Curtis~McCully$^{\ref{af:LCO}}$,
Megan~Newsome$^{\ref{af:UT}}$,\\
Estefania~Padilla~Gonzalez$^{\ref{af:JHU}}$,
\&
Giacomo~Terreran$^{\ref{af:Adler}}$
\\
\normalsize{*Corresponding author. Email: dhiramatsu@ufl.edu}
}

\begin{document}

\maketitle

\begin{affiliations}
 \item Department of Astronomy, University of Florida, Bryant Space Science Center, Gainesville, FL 32611-2055, USA\label{af:UF}
 \item Center for Astrophysics \textbar{} Harvard \& Smithsonian, 60 Garden Street, Cambridge, MA 02138-1516, USA\label{af:CfA}
 \item The NSF AI Institute for Artificial Intelligence and Fundamental Interactions, USA\label{af:IAIFI}
 \item TAPIR, Mailcode 350-17, California Institute of Technology, Pasadena, CA 91125-0001, USA\label{af:TAPIR}
 \item Research Center for the Early Universe, School of Science, The University of Tokyo, 7-3-1 Hongo, Bunkyo-ku, Tokyo 113-0033, Japan\label{af:RESCEU}
 \item Department of Astronomy, The University of Texas at Austin, 2515 Speedway, Stop C1400, Austin, TX 78712, USA\label{af:UT}
 \item National Astronomical Observatory of Japan, National Institutes of Natural Sciences, 2-21-1 Osawa, Mitaka, Tokyo 181-8588, Japan\label{af:NAOJ}
 \item Graduate Institute for Advanced Studies, SOKENDAI, 2-21-1 Osawa, Mitaka, Tokyo 181-8588, Japan\label{af:SOKENDAI}
 \item School of Physics and Astronomy, Monash University, Clayton, VIC 3800, Australia\label{af:Monash}
 \item Las Cumbres Observatory, 6740 Cortona Drive, Suite 102, Goleta, CA 93117-5575, USA\label{af:LCO}
 \item Department of Physics, University of California, Santa Barbara, CA 93106-9530, USA\label{af:UCSB}
 \item Department of Astronomy, University of California, Berkeley, CA 94720-3411, USA\label{af:UCB}
 \item Physics Department, Tsinghua University, Beijing 100084, China\label{af:Tsinghua}
 \item Steward Observatory, University of Arizona, 933 North Cherry Avenue, Tucson, AZ 85721-0065, USA\label{af:UA}
 \item LSSTC Catalyst Fellow\label{af:Catalyst}
 \item Department of Physics and Astromomy, Johns Hopkins University, Baltimore, MD 21210, USA\label{af:JHU}
 \item Adler Planetarium, 1300 South DuSable Lake Shore
Drive, Chicago, IL 60605, USA\label{af:Adler}
\end{affiliations}



\clearpage

\begin{abstract}

Stars in the initial and carbon-oxygen core mass ranges\cite{Heger2002ApJ...567..532H,Heger2003ApJ...591..288H,Takahashi2018ApJ...857..111T,Renzo2024arXiv240716113R} of $\sim140-260$ and $50-130\,$M$_\odot$, respectively, with low metallicity\cite{Langer2007A&A...475L..19L,Yusof2013MNRAS.433.1114Y} are predicted to experience copious electron-positron pair production in their cores\cite{Barkat1967PhRvL..18..379B,Rakavy1967ApJ...148..803R}, leading to a runaway thermonuclear explosion that obliterates the entire star in a luminous and long-duration pair-instability supernova explosion\cite{Kasen2011ApJ...734..102K,Dessart2013MNRAS.428.3227D,Kozyreva2014A&A...565A..70K,Gilmer2017ApJ...846..100G}. Some previous supernovae\cite{Terreran2017NatAs...1..713T,Kozyreva2018MNRAS.479.3106K} have been interpreted in this context but lack the full range of predicted properties. Here, we report detailed observations and modeling of the hydrogen-rich supernova 2023vbw, which exploded in a low-metallicity ($\sim0.1$\,Z$_\odot$) environment in a dwarf star-forming galaxy at a redshift of $0.088$. Its light curve exhibits a luminous ($1.6\times10^{43}$\,erg\,s$^{-1}$) and long-duration ($190$ days) main peak, resulting in a total radiated energy of $3\times10^{50}$\,erg, more than an order of magnitude greater than canonical core-collapse supernovae. Semi-analytical light-curve modeling\cite{Pumo2025MNRAS.538..223P} yields a blue supergiant-like progenitor with an ejecta mass of $170-350$\,M$_\odot$, radioactive nickel mass of $1.2-1.6$\,M$_\odot$, and explosion energy of $(6-13)\times10^{52}$\,erg, well matched by pair-instability models. The early and late-phase light curve and spectra also show evidence for interaction of the supernova ejecta with an aspherical circumstellar medium. 
Discoveries of numerous such events with the upcoming Rubin Observatory and \textit{Roman Space Telescope} will shed light on the deaths of the most massive stars in the Universe. 

\end{abstract}



\begin{figure}
\centering
\includegraphics[width=0.99\textwidth]{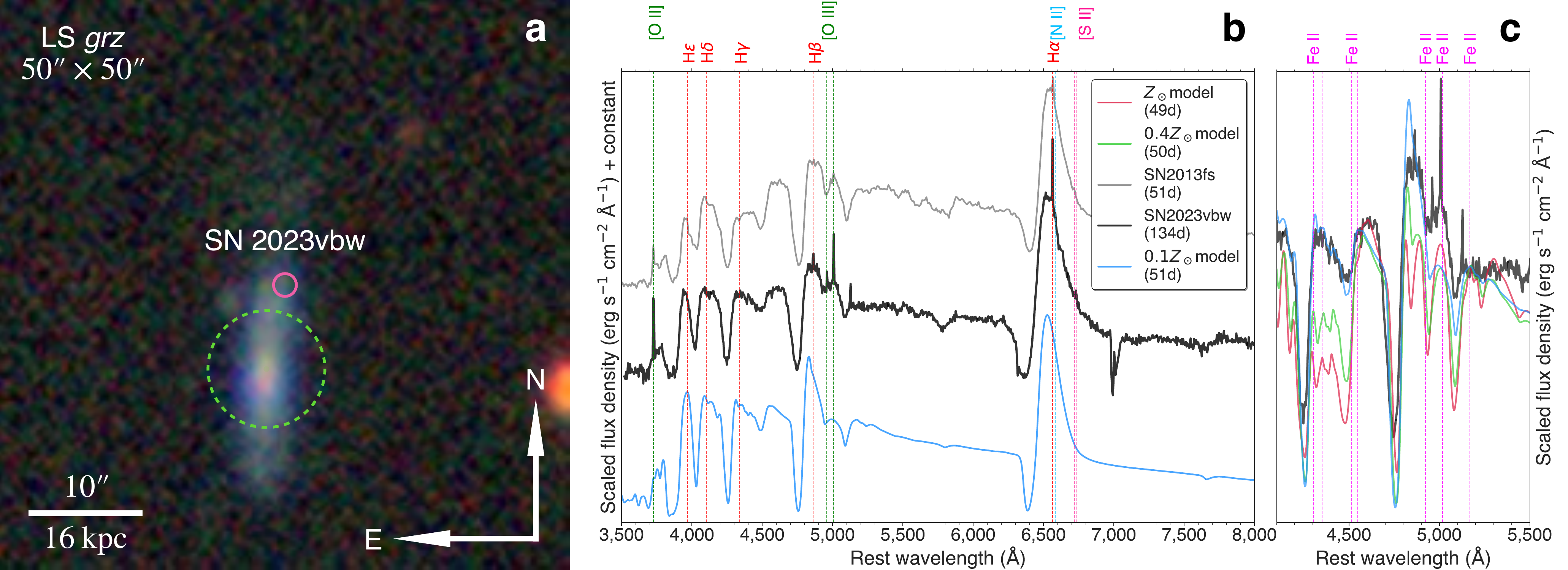}
\caption{
\textbf{Supernova environment and spectral classification.}
\textbf{a,} The location of SN~2023vbw (J2000 $\text{R.A.}=08^{\text{hr}}15^{\text{m}}04^{\text{s}}.356$, $\text{decl.}=+57^{\circ}10'55".60$; magenta circle) in the outskirts of its dwarf host galaxy (green circle; see Extended Data Fig.~\ref{EDfig:SED} for the integrated SED), which has a Kron radius of $5{\farcs}1=8.2$ kpc. Image from the DESI Legacy Imaging Surveys\cite{DESILIS}.
\textbf{b, c,} Spectral comparison of our classification spectrum of SN~2023vbw at 134\,d with the best-match SN~II~2013fs at 51\,d from cross-correlation\cite{Goldwasser2022TNSAN.191....1G} as well as theoretical SN~II models\cite{Dessart2013MNRAS.433.1745D} with varying metallicities. The broad SN hydrogen Balmer and Fe~{\sc ii} lines, as well as the narrow host emission lines, are marked at their rest wavelengths. 
The Fe~{\sc ii} absorption lines are weaker in SN~2023vbw than in the $0.1\,{\rm Z}_\odot$ model (\textbf{c}), indicating an even lower metallicity for the progenitor star.
\label{fig:class}
}
\end{figure}

On 2023 October 12.5 (UTC dates are used throughout), the Zwicky Transient Facility (ZTF) discovered\cite{Forster2023TNSTR2596....1F} the transient AT~2023vbw with an $r$-band magnitude of $19.9$ in the outskirts of its dwarf host galaxy (Fig.~\ref{fig:class}). Using the last nondetection and first prediscovery detection from ZTF forced photometry (Methods), we estimate an epoch of first light of 2023 October 10.0\,$\pm$\,1.4 and use it as a reference epoch for all phases. 
Following a tentative spectroscopic classification\cite{Perez2024TNSCR.569....1P} as a hydrogen-rich Type~II supernova (SN~II), we confirmed the classification with a higher signal-to-noise ratio spectrum at 134 days (Fig.~\ref{fig:class}) and measured a precise redshift of 0.0879 from narrow host-galaxy emission lines (Methods), corresponding to a luminosity distance\cite{Riess2022ApJ...934L...7R} of 396 Mpc. Given the lack of Na~{\sc i}~D absorption at the host redshift, we correct all data only for the Milky Way extinction\cite{Schlafly2011ApJ...737..103S} ($A_V=0.126$ mag) assuming the reddening law\cite{Cardelli1989ApJ...345..245C} with $R_V=3.1$.

Our classification spectrum of SN~2023vbw is characterized by broad Balmer P~Cygni profiles without narrow emission components, typical of SNe~II\cite{Filippenko1997ARA&A..35..309F}. However, the best match from spectral cross-correlation\cite{Goldwasser2022TNSAN.191....1G} is with an SN phase $\sim 80$ days earlier, indicating the unusually slow evolution of SN~2023vbw. A comparison to theoretical spectral models\cite{Dessart2013MNRAS.433.1745D} reveals a similar phase offset, and a best match to a low-metallicity ($0.1\,{\rm Z}_\odot$ with ${\rm Z}_\odot=0.02$) model based on weak Fe~{\sc ii} absorption. 
This is consistent with a gas-phase metallicity of $\sim0.3\,{\rm Z}_\odot$ inferred from host emission lines at the SN location (Methods). The integrated host spectral energy distribution (SED) also reveals overall low gas-phase ($\sim 0.2\,{\rm Z}_\odot$) and stellar ($\sim 0.08\,{\rm Z}_\odot$) metallicities, active star formation ($\sim 0.2\,{\rm M}_\odot\,{\rm yr}^{-1}$), and low stellar mass ($\sim 10^{9.3}\,{\rm M}_\odot$) (Extended Data Fig.~\ref{EDfig:SED}; Methods).

\begin{figure}
\centering
\includegraphics[width=0.99\textwidth]{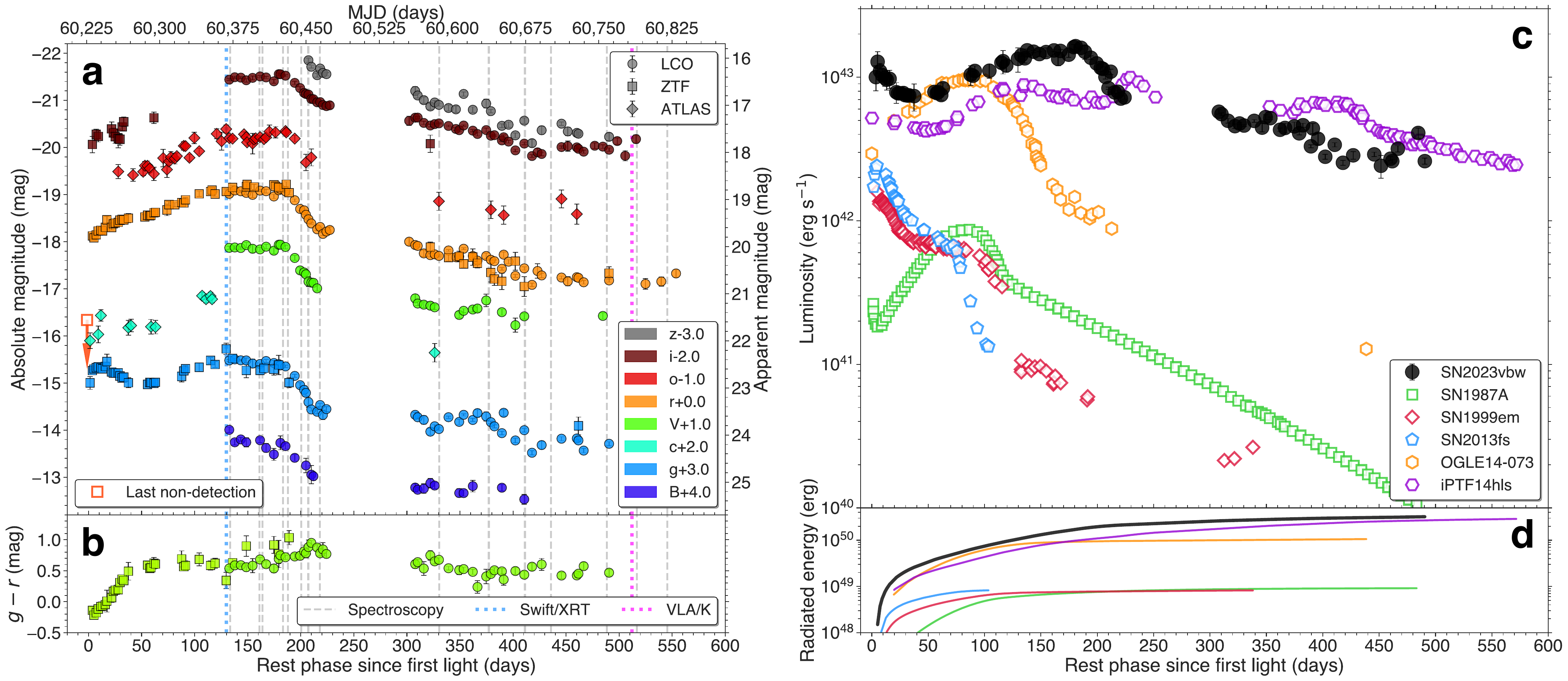}
\caption{
\textbf{Optical and bolometric light curves.}
\textbf{a, b,} The multiband and $g-r$ colour  curves of SN~2023vbw. Error bars denote $1\sigma$ uncertainties. The epochs  of spectral (Fig.~\ref{fig:spec}) and multiwavelength (Methods) observations are marked by the vertical dashed and dotted lines, respectively. SN~2023vbw displays four light-curve phases: initial cooling with blue colour ($\lesssim40$ d), steady rise with constant colour ($\sim40-190$ d), rapid decline with red colour ($\sim190-230$ d), and extended tail with slightly bluer colour than the main peak ($\gtrsim230$ d). The peak absolute magnitude in the $r$ band is $\sim -19.2$.
\textbf{c, d,} Comparison of the bolometric light curve and cumulative radiated energy of SN~2023vbw with prototypical SNe~II\cite{Valenti2016MNRAS.459.3939V} 1999em and 2013fs, SN~II~1987A\cite{Catchpole1989MNRAS.237P..55C}, the energetic SN~1987A-like OGLE14-073\cite{Terreran2017NatAs...1..713T}, and the peculiar SN~II iPTF14hls\cite{Arcavi2017Natur.551..210A,Sollerman2019A&A...621A..30S}. The light-curve morphology of SN~2023vbw resembles that of SN~1987A and OGLE14-073, but with higher luminosity and longer timescale, resulting in a larger radiated energy of $\sim 3\times10^{50}$\,erg. The luminosity and timescale of the tail fluctuations are comparable to those of iPTF14hls.
\label{fig:LC}
}
\end{figure}

Consistent with the slow spectral evolution, the optical light curves (Fig.~\ref{fig:LC}) exhibit an extended cooling phase ($\sim 40$ days) with blue colour (evolving from $g-r\approx-0.2$ to $0.3$ mag), followed by a long steady rise to a bright peak of $M_r=-19.2$ mag at $\sim 190$ days with nearly constant colour ($g-r\approx0.6$ mag). The light curves then rapidly decline by $1$\,mag in $\sim40$ days with red colour (up to $g-r\approx0.9$ mag) and subsequently settle on an extended tail with a mean decline rate of $\sim0.005$\,mag\,day$^{-1}$, slower than the $^{56}$Co decay rate of $\sim0.01$\,mag\,day$^{-1}$, with mild fluctuations and slightly bluer colour ($g-r\approx0.5$ mag) than the main peak. 

Blackbody fits to the multi-epoch optical SEDs (Methods) reveal rapid temperature and radius evolution in the initial cooling phase, from $\sim14{,}000$ to $7{,}000$\,K and $\sim 0.6\times 10^{15}$ to $2\times10^{15}$\,cm, respectively, followed by a constant temperature ($\sim5{,}000$\,K) and more gradual increase in radius (up to $\sim6\times10^{15}$\,cm) during the rise to peak (Extended Data Fig.~\ref{EDfig:BB}). During the decline and tail phases, the photospheric radius begins to recede (down to $\sim2\times10^{15}$\,cm, matching the end of the initial cooling phase), while the temperature slightly rises (up to $\sim6{,}000$\,K).

A comparison of the bolometric light curve with prototypical SNe~II\cite{Valenti2016MNRAS.459.3939V} (e.g., SNe~1999em and 2013fs) from red supergiant (RSG) progenitors highlights the unusual nature of SN~2023vbw (Fig.~\ref{fig:LC}), which more closely resembles the light-curve morphology of SN~II~1987A\cite{Catchpole1989MNRAS.237P..55C} from a compact blue supergiant (BSG) progenitor, but with significantly larger luminosity and timescale. This results in a total radiated energy of $\sim3\times10^{50}$\,erg, more than an order of magnitude larger than normal SNe~II, which radiate $\sim10^{49}\,{\rm erg}$ ($\sim 1\%$ of their explosion energy). SN~2023vbw even exceeds previously known very energetic SNe~II (e.g., the SN~1987A-like OGLE14-073\cite{Terreran2017NatAs...1..713T,Kozyreva2018MNRAS.479.3106K} and the peculiar iPTF14hls\cite{Arcavi2017Natur.551..210A,Sollerman2019A&A...621A..30S}), likely requiring a higher explosion energy than achievable with the neutrino-driven mechanism\cite{Sukhbold2016ApJ...821...38S} ($\lesssim2\times10^{51}$ erg) for iron core-collapse SNe (CCSNe). The late-time fluctuations of SN~2023vbw ($\gtrsim330$ days) have comparable luminosity and timescale to those of iPTF14hls.

\begin{figure}
\centering
\includegraphics[width=0.99\textwidth]{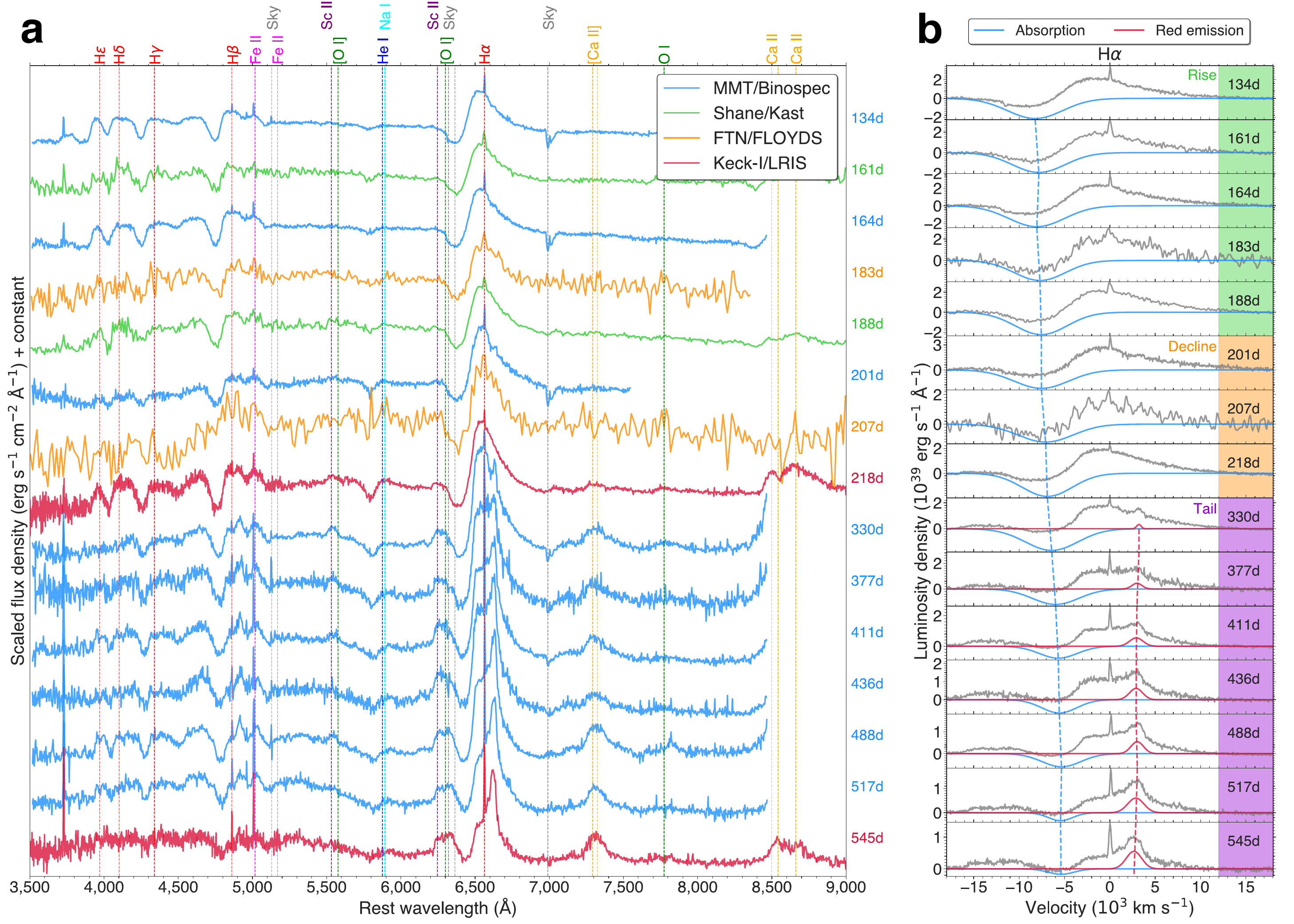}
\caption{
\textbf{Optical spectral time series.}
\textbf{a, b,} The optical spectra and H$\alpha$ line profiles of SN~2023vbw in the phase range of 134 to 545 d. Prominent SN spectral features of the hydrogen Balmer series and various metal lines commonly observed in SNe II are marked at their rest wavelengths. The spectra remain almost unchanged with a constant absorption velocity during the light-curve rise at $\lesssim 190$ d (absorption minimum marked with the blue dashed line; \textbf{b}), begin to display forbidden lines (i.e., [O~{\sc i}] and [Ca~{\sc ii}]) and decreasing velocity during the light-curve decline at $\sim 200-220$ d, and finally exhibit a multicomponent line profile with an increasing redshifted emission component (emission peak marked with the red dashed line; \textbf{b}) during the light-curve tail at $\gtrsim 330$ d.   
\label{fig:spec}
 }
\end{figure}

The various light-curve phases (i.e., rise, decline, and tail) are accompanied by distinctive spectral evolution (Fig.~\ref{fig:spec}). During the light-curve rise, the typical SN~II features remain steady, with a roughly constant velocity ($\sim8{,}000$ and $4{,}500$\,km\,s$^{-1}$ for H$\alpha$ and Fe~{\sc ii}\,$\lambda 5{,}169$, respectively; Extended Data Fig.~\ref{EDfig:vel}). This indicates that the photosphere remains at the same mass coordinate, which is physically expanding with a constant temperature (Extended Data Fig.~\ref{EDfig:BB}), requiring a large continuous heating source, unlike typical SNe~II, but similar to SN~1987A and iPTF14hls. 
Forbidden lines emerge during the light-curve decline, gradually shifting into the optically thin phase with a receding photosphere and decreasing velocity (Extended Data Figs.~\ref{EDfig:BB}\,\&\,\ref{EDfig:vel}). During the light-curve tail, the H$\alpha$ line evolves to exhibit a multicomponent profile with prominent red emission at a velocity offset of $\sim2{,}800$\,km\,s$^{-1}$ (similarly seen in other hydrogen lines; Extended Data Fig.~\ref{EDfig:nir}). This indicates asymmetry in the photosphere, likely revealing previously hidden shock interaction with disc-like circumstellar material (CSM) viewed off-axis\cite{Smith2015MNRAS.449.1876S,Kurf2020A&A...642A.214K} (Methods), as similarly observed in iPTF14hls\cite{Andrews2018MNRAS.477...74A,Sollerman2019A&A...621A..30S}, which provides the rise in temperature inferred from the light curve (Extended Data Fig.~\ref{EDfig:BB}) and responsible for the persistent continuum below $5$,$500\,\Angstrom$ in the spectra (Extended Data Fig.~\ref{EDfig:nir}). The lack of the narrow absorption features from unshocked CSM places an upper limit on its velocity of $\lesssim170$\,km\,s$^{-1}$ (Methods). The coincidence of the blackbody radii at the tail phase ($\gtrsim400$ days) and the end of the intial cooling phase ($\sim40$ days) indicates that the early phase may also be powered by CSM interaction (until the more spherical SN ejecta become dominant during the light-curve rise and peak).

\begin{figure}
\centering
\includegraphics[width=0.99\textwidth]{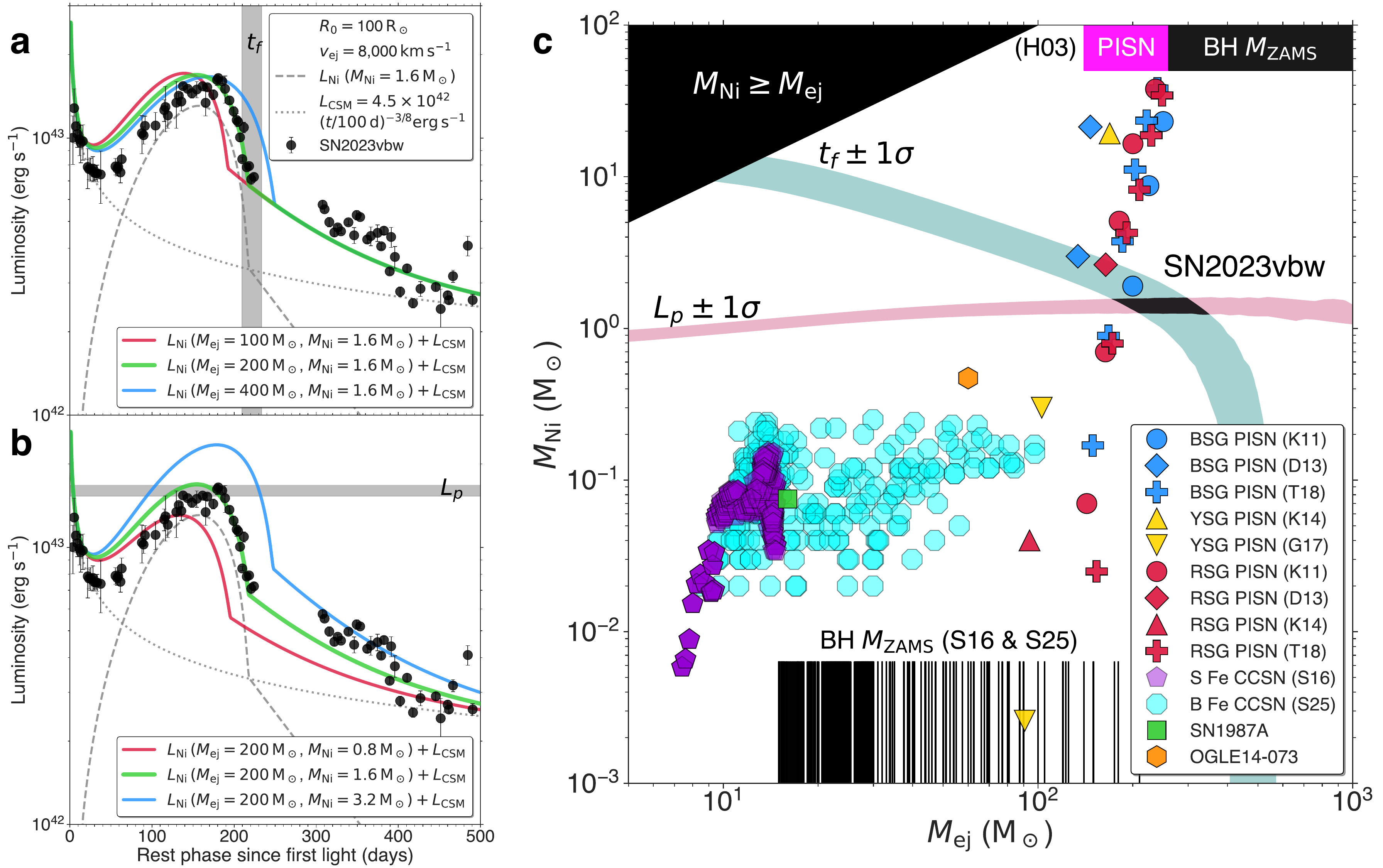}
\caption{
\textbf{Model fits and parameter constraints.}
\textbf{a, b,} The $^{56}$Ni+CSM light-curve models (solid lines) with varying ejecta masses ($M_{\rm ej}$; \textbf{a}) and $^{56}$Ni masses ($M_{\rm Ni}$; \textbf{b}). The $^{56}$Ni and CSM contributions are shown as grey dashed and dotted lines, respectively. The recombination timescale ($t_f=215^{+19}_{-6}$ days, i.e., the end of optically thick phase; vertical grey bar) and peak luminosity ($L_p=(1.60\pm0.07)\times10^{43}$\,erg\,s$^{-1}$; horizontal grey bar) are measured from the Gaussian-Process light curve (Extended Data Fig.~\ref{EDfig:BB}; Methods) and used for determining the model parameters. 
\textbf{c,} Allowed parameter space of $M_{\rm ej}$ and $M_{\rm Ni}$ based on the $t_f\pm1\sigma$ and $L_p\pm1\sigma$ ranges, with the intersection region marking the allowed parameter ranges for SN~2023vbw ($M_{\rm ej}\approx170-350\,{\rm M}_\odot$ and $M_{\rm Ni}\approx1.2-1.6\,{\rm M}_\odot$; dark grey shade).  Also shown are the results from numerical PISN models (K11\cite{Kasen2011ApJ...734..102K}, D13\cite{Dessart2013MNRAS.428.3227D}, K14\cite{Kozyreva2014A&A...565A..70K}, G17\cite{Gilmer2017ApJ...846..100G}, and T18\cite{Takahashi2018ApJ...857..111T}) from BSG, yellow supergiant (YSG), and RSG-like progenitors with the zero-age main sequence masses of $M_{\rm ZAMS}=150-260\,{\rm M}_\odot$ at $\leq10^{-3}\,{\rm Z}_\odot$, numerical hydrogen-rich iron CCSN models from single stellar evolution (S16\cite{Sukhbold2016ApJ...821...38S}) with $M_{\rm ZAMS}=9-120\,{\rm M}_\odot$ and binary stellar evolution (S25\cite{Schneider2025A&A...700A.253S}) with $M_{\rm ZAMS}=11-70\,{\rm M}_\odot$ both at ${\rm Z}_\odot$, and the observed SNe~II 1987A\cite{Orlando2015ApJ...810..168O,Pumo2025MNRAS.538..223P} and OGLE14-073\cite{Terreran2017NatAs...1..713T,Pumo2025MNRAS.538..223P} at $\sim0.5\,{\rm Z}_\odot$.
The bottom black vertical lines show $M_{\rm ZAMS}$ of failed CCSN (i.e., direct black hole collapse) models from S16 and S25, while the top-right magenta and black shaded regions show $M_{\rm ZAMS}$ ranges of PISN and direct black hole collapse models, respectively, from H03\cite{Heger2003ApJ...591..288H}.
Note that $M_{\rm ej} < M_{\rm ZAMS}$ due to the metallicity-dependent progenitor mass loss. 
The allowed parameter space of SN~2023vbw resides far from the iron CCSN regime, but well within the PISN regime (see also Extended Data Fig.~\ref{EDfig:PISN_mod} for the light-curve and velocity comparisons with the nearest neighbor PISN models).
\label{fig:mod}
}
\end{figure}

Motivated by the SN~1987A-like light-curve morphology (Fig.~\ref{fig:LC}) and velocity evolution (Extended Data Fig.~\ref{EDfig:vel}) of SN~2023vbw, we use a semi-analytical light-curve model\cite{Pumo2025MNRAS.538..223P} (Methods) that calculates the bolometric luminosity by tracking hydrogen recombination in the ejecta and energy deposition by $^{56}$Ni. We assume an outermost scale velocity of $v_{\rm sc}=8{,}000$\,km\,s$^{-1}$ from the H$\alpha$ velocity (i.e., proxy for the fastest ejecta velocity in homologous phase), relating the ejecta kinetic energy and mass as $E_{\rm ej}=0.3M_{\rm ej}v_{\rm sc}^2$, and a BSG-like progenitor radius\cite{Kasen2011ApJ...734..102K,Dessart2013MNRAS.428.3227D,Takahashi2018ApJ...857..111T} of $R_0=100\,{\rm R}_\odot$ (Methods; see Extended Data Fig.~\ref{EDfig:mod_var} for the light-curve model dependence on $R_0$, preferring a compact progenitor).  The main peak is well fit with a narrow region of the parameter space: ejecta and $^{56}$Ni masses of $M_{\rm ej}\approx170-350\,{\rm M}_\odot$ and $M_{\rm Ni}\approx 1.2-1.6\,{\rm M}_\odot$, respectively (Fig.~\ref{fig:mod}). These masses, and the resultant kinetic energy of $E_{\rm ej}\sim(6-13)\times10^{52}$\,erg, exceed the values for ordinary iron CCSNe\cite{Sukhbold2016ApJ...821...38S,Schneider2025A&A...700A.253S} by more than an order of magnitude (e.g., $M_{\rm ej}\approx16\,{\rm M}_\odot$ and $M_{\rm Ni}\approx0.075\,{\rm M}_\odot$ with $\sim 1.3\times10^{51}$\,erg for SN~1987A\cite{Orlando2015ApJ...810..168O,Pumo2025MNRAS.538..223P}), and significantly exceed even the extreme SN~II OGLE14-073\cite{Terreran2017NatAs...1..713T,Pumo2025MNRAS.538..223P} ($M_{\rm ej}\approx60\,{\rm M}_\odot$ and $M_{\rm Ni}\approx 0.47\,{\rm M}_\odot$ with $\sim 1.2\times10^{52}$\,erg).


The initial extended cooling phase, as well as the tail-phase excess relative to $^{56}$Co decay, can both be fit by a single power-law component with time $t^{-3/8}$ (Fig.~\ref{fig:mod}), whose evolution is consistent with interaction of the SN ejecta with dense CSM having a wind-like density profile with mass-loss rate of $\sim 0.05\,{\rm M}_\odot\,{\rm yr}^{-1} (v_{\rm CSM}/170\,{\rm km\,s}^{-1}) (v_{\rm sc}/8{,}000\,{\rm km\,s}^{-1})^{-3}$, where $v_{\rm CSM}$ is the CSM velocity, scaled to its upper limit (Methods). The CSM interaction is also evident by the emergence of strong double-peaked spectral features during the tail phase (Fig.~\ref{fig:spec} \& Extended Data Fig.~\ref{EDfig:nir}), which points to a disc-like CSM configuration viewed off-axis.


The observed and modeled properties of SN~2023vbw are well matched to the regime of pair-instability supernova (PISN) predictions\cite{Heger2002ApJ...567..532H,Heger2003ApJ...591..288H,Takahashi2018ApJ...857..111T,Kasen2011ApJ...734..102K,Dessart2013MNRAS.428.3227D} (Fig.~\ref{fig:mod}; see also Extended Data Fig.~\ref{EDfig:PISN_mod} for the compatible light-curve and velocity evolution with the nearest neighbor PISN numerical models from BSG progenitors). The low, but nonzero, metallicity ($\sim 0.1\,{\rm Z}_\odot$) inferred from the SN itself and its environment (Fig.~\ref{fig:class} \& Extended Data Fig.~\ref{EDfig:SED}) is also predicted for PISNe\cite{Langer2007A&A...475L..19L,Yusof2013MNRAS.433.1114Y}. 
Although the progenitor formation channel to retain the sufficient hydrogen-rich envelope within a small BSG-like radius is not well understood, one possible channel is a binary merger\cite{Podsiadlowski1992ApJ...391..246P,Vigna2019ApJ...876L..29V}, which may also explain the presence of the dense disc-like CSM\cite{Morris2007Sci...315.1103M} (Methods). 
Alternative iron CCSN scenarios with an additional power source, such as a central magnetar\cite{Kasen2010ApJ...717..245K} or black hole accretion\cite{Dexter2013}, yield comparable $M_{\rm ej}$ to fit the light-curve main peak due to photon diffusion (Methods); however, magnetar formation is not naturally expected in the high progenitor mass range\cite{Heger2003ApJ...591..288H,Takahashi2018ApJ...857..111T}, and sustained accretion ($\gtrsim 100$ days) is likely implausible given the required large fallback mass and energy deposition\cite{Dexter2013} (Methods).


Thanks to its relatively low redshift, SN~2023vbw remains sufficiently bright for continued multiwavelength observations that will reveal its progenitor mass-loss history and explosive nucleosynthesis. With declining optical depth, synchrotron emission from CSM interaction is expected to become detectable in the X-rays and radio (Methods), while strong nebular emission lines from iron-group elements are expected to emerge in the optical and near-infrared\cite{Dessart2013MNRAS.428.3227D,Jerkstrand2016MNRAS.455.3207J}. Scaling from our discovery of SN~2023vbw in current surveys, we expect that upcoming surveys with a significant increase in search volume and redshift reach, namely the Vera C. Rubin Observatory and \textit{Nancy Grace Roman Space Telescope}, will find tens to hundreds of such events, shedding light on the environment, formation, evolution, and death of the most massive stars in the Universe\cite{Heger2002ApJ...567..532H,Heger2003ApJ...591..288H,Takahashi2018ApJ...857..111T,Renzo2024arXiv240716113R}, as well as on their wide-ranging implications from the mass limits for black hole formation\cite{Woosley2021ApJ...912L..31W} to the unique fingerprints on early chemical evolution\cite{Kobayashi2025arXiv250620436K}.

%
%

%




 
\begin{addendum}

\item[Acknowledgements]

We are grateful to Masaomi Tanaka, Kazumi Kashiyama, Shigeo S. Kimura, Alex Gagliano, Jared A. Goldberg, Anthony L. Piro, Wynn V. Jacobson-Gal\'{a}n, David Vartanyan, Lars Bildsten, and Iair Arcavi for useful discussions, and to Sean Moran and Benjamin Weiner for scheduling the MMT observations.

D.H. is supported by NASA grants HST-GO-17770.002, JWST-GO-12468.001, and JWST-GO-09964.001.
This work is supported by the U.S. National Science Foundation (NSF) under Cooperative Agreement PHY-2019786 (The NSF AI Institute for Artificial Intelligence and Fundamental Interactions, \url{http://iaifi.org/}).
D.T. is supported by the Sherman Fairchild Postdoctoral Fellowship at Caltech and the Institute for Theory and Computation Fellowship at CfA.
A.V.F.’s group at UC Berkeley is grateful for financial assistance from the Christopher R. Redlich Fund, Gary and Cynthia Bengier, Clark and Sharon Winslow, Alan Eustace and Kathy Kwan (W.Z. is a Bengier-Winslow-Eustace Specialist in Astronomy), William Draper, Timothy and Melissa Draper, Briggs and Kathleen Wood, Sanford Robertson (T.G.B. is a Draper-Wood-Robertson Specialist in Astronomy), and many other donors.
Y.Y.'s research is now partially supported by the Tsinghua University Dushi Program, and was previously supported through a Benoziyo Prize Postdoctoral Fellowship and the Bengier-Winslow-Robertson Fellowship.
The LCO team is supported by NSF grants AST-1911225 and AST-1911151.
This publication was made possible through the support of an LSSTC Catalyst Fellowship to K.A.B., funded through grant 62192 from the John Templeton Foundation to LSST Corporation. The opinions expressed in this publication are those of the authors and do not necessarily reflect the views of LSSTC or the John Templeton Foundation.

Observations reported here were obtained at the MMT Observatory, a joint facility of the Smithsonian Institution and the University of Arizona.

This work makes use of observations from the Las Cumbres Observatory network. This paper is based in part on observations made with the MuSCAT3 instrument, developed by the Astrobiology Center and under financial support by JSPS KAKENHI (grant No. JP18H05439) and JST PRESTO (grant No. JPMJPR1775), at Faulkes Telescope North on Maui, HI, operated by the Las Cumbres Observatory. The authors wish to recognize and acknowledge the very significant cultural role and reverence that the summit of Haleakal\={a} has always had within the indigenous Hawaiian community. We are most fortunate to have the opportunity to conduct observations from the mountain. 

Some of the data presented herein were obtained at the W. M. Keck
Observatory, which is operated as a scientific partnership among the
California Institute of Technology, the University of California, and
NASA; the observatory was made possible by the generous financial
support of the W. M. Keck Foundation.
A major upgrade of the Kast spectrograph on the Shane 3 m telescope at Lick Observatory, led by Brad Holden, was made possible through generous gifts from the Heising-Simons Foundation, William and Marina Kast, and the University of California Observatories.  Research at Lick Observatory is partially supported by a generous gift from Google. 

Based on observations obtained at the international Gemini Observatory, a program of NSF NOIRLab, which is managed by the Association of Universities for Research in Astronomy (AURA) under a cooperative agreement with the U.S. National Science Foundation on behalf of the Gemini Observatory partnership: the U.S. National Science Foundation (United States), National Research Council (Canada), Agencia Nacional de Investigaci\'{o}n y Desarrollo (Chile), Ministerio de Ciencia, Tecnolog\'{i}a e Innovaci\'{o}n (Argentina), Minist\'{e}rio da Ci\^{e}ncia, Tecnologia, Inova\c{c}\~{o}es e Comunica\c{c}\~{o}es (Brazil), and Korea Astronomy and Space Science Institute (Republic of Korea).

This work has made use of data from the Zwicky Transient Facility (ZTF).  ZTF is supported by NSF grant No. AST-1440341 and a collaboration including Caltech, IPAC, the Weizmann Institute for Science, the Oskar Klein Center at Stockholm University, the University of Maryland, the University of Washington, Deutsches Elektronen-Synchrotron and Humboldt University, Los Alamos National Laboratories, the TANGO Consortium of Taiwan, the University of Wisconsin at Milwaukee, and Lawrence Berkeley National Laboratories. Operations are conducted by COO, IPAC, and UW. The ZTF forced-photometry service was funded under the Heising-Simons Foundation grant No. 12540303 (PI: Graham).

This work has made use of data from the Asteroid Terrestrial-impact Last Alert System (ATLAS) project. ATLAS is primarily funded to search for near-Earth asteroids through NASA grant Nos. NN12AR55G, 80NSSC18K0284, and 80NSSC18K1575; byproducts of the NEO search include images and catalogs from the survey area. This work was partially funded by Kepler/K2 grant No. J1944/80NSSC19K0112 and HST grant No. GO-15889, and STFC grant Nos. ST/T000198/1 and ST/S006109/1. The ATLAS science products have been made possible through the contributions of the University of Hawaii Institute for Astronomy, the Queen’s University Belfast, the Space Telescope Science Institute, the South African Astronomical Observatory, and The Millennium Institute of Astrophysics (MAS), Chile.

The Legacy Surveys consist of three individual and complementary projects: the Dark Energy Camera Legacy Survey (DECaLS; Proposal ID \#2014B-0404; PIs: David Schlegel and Arjun Dey), the Beijing-Arizona Sky Survey (BASS; NOAO Prop. ID \#2015A-0801; PIs: Zhou Xu and Xiaohui Fan), and the Mayall z-band Legacy Survey (MzLS; Prop. ID \#2016A-0453; PI: Arjun Dey). DECaLS, BASS and MzLS together include data obtained, respectively, at the Blanco telescope, Cerro Tololo Inter-American Observatory, NSF’s NOIRLab; the Bok telescope, Steward Observatory, University of Arizona; and the Mayall telescope, Kitt Peak National Observatory, NOIRLab. The Legacy Surveys project is honored to be permitted to conduct astronomical research on Iolkam Du’ag (Kitt Peak), a mountain with particular significance to the Tohono O’odham Nation.

This work made use of data supplied by the UK Swift Science Data Centre at the University of Leicester.

The National Radio Astronomy Observatory and Green Bank Observatory are facilities of the U.S. NSF operated under cooperative agreement by Associated Universities, Inc.

This research has made use of the NASA Astrophysics Data System (ADS), the NASA/IPAC Extragalactic Database (NED), NASA/IPAC Infrared Science Archive (IRSA, which is funded by NASA and operated by the California Institute of Technology), and IRAF (which is distributed by the National Optical Astronomy Observatory, NOAO, operated by the Association of Universities for Research in Astronomy, AURA, Inc., under cooperative agreement with the NSF).

TNS is supported by funding from the Weizmann Institute of Science, as well as grants from the Israeli Institute for Advanced Studies and the European Union via ERC grant No. 725161.

\item[Author Contributions]
Daichi Hiramatsu initiated the study, organized the follow-up observations, processed the Las Cumbres data, performed the analysis, and led the writing of the manuscript.
Edo Berger assisted with the scientific interpretation and writing the manuscript. 
Daichi Tsuna produced the light-curve models and assisted with their interpretation and writing the manuscript.
Sebastian Gomez, Harsh Kumar, and Peter K. Blanchard assisted in obtaining the MMT spectra; Gomez and Kumar reduced them. Kumar also obtained and reduced the Gemini spectra.
Anya E. Nugent performed the host galaxy SED fit and assisted with its interpretation and writing the manuscript.
Walter W. Golay facilitated the submission of VLA observations, performed the data reduction and analysis, and assisted with manuscript composition.
Takashi J. Moriya assisted with theoretical PISN progenitor and light curve interpretations.
D. Andrew Howell is the principal investigator of the Las Cumbres Observatory Global Supernova Project through which all of the Las Cumbres data were obtained; he also assisted with the data interpretation.
Alexei V. Filippenko, Thomas G. Brink, WeiKang Zheng, and Yi Yang obtained the Lick and/or Keck spectra; Brink reduced them. Filippenko also edited the manuscript.
Moira Andrews, K. Azalee Bostroem, Joseph Farah, Curtis McCully, Megan Newsome, Estefania Padilla Gonzalez, and Giacomo Terreran assisted in obtaining and reducing the Las Cumbres data.
 
\item[Author Information] The authors declare that they have no competing financial interests. Correspondence and requests for materials should be addressed to D.~Hiramatsu (dhiramatsu@ufl.edu).


\end{addendum}


\clearpage

\newpage

\newpage

\newpage

\clearpage

\begin{methods}

\subsection{Follow-up photometry.}
\label{sec:imaging}

Through the Global Supernova Project (GSP), we obtained Las Cumbres Observatory (LCO\cite{Brown2013PASP..125.1031B}) $BVgri$-band imaging with the Sinistro cameras on the network of 1\,m telescopes at the McDonald Observatory (Texas, USA) and the Teide Observatory (Canary Islands, Spain), as well as $griz$-band imaging with the Multicolor Simultaneous Camera for studying Atmospheres of Transiting exoplanets 3 (MuSCAT3\cite{Narita2020SPIE11447E..5KN}) on the 2\,m Faulkes Telescope North (FTN) at the Haleakal\={a} Observatory (Hawai`i, USA), from 2024 March 2 to 2025 June 3 (133 to 554 days after first light). 
LCO photometry was performed with point-spread-function (PSF) fitting using the PyRAF\cite{PyRAF2012ascl.soft07011S}-based \texttt{lcogtsnpipe}\cite{Valenti2016MNRAS.459.3939V} pipepline. $BV$- and $griz$-band data were calibrated to Vega\cite{Johnson1953ApJ...117..313J} and AB\cite{Oke1983ApJ...266..713O} magnitudes, respectively, with the AAVSO Photometric All Sky Survey\cite{Henden2016yCat.2336....0H} and Sloan Digital Sky Survey (SDSS\cite{Albareti2017ApJS..233...25A}).

Additionally, we collected public photometry from ZTF\cite{Bellm2019PASP..131a8002B, Graham2019PASP..131g8001G} and the Asteroid Terrestrial-impact Last Alert System (ATLAS\cite{Tonry2018PASP..130f4505T, Smith2020PASP..132h5002S}) up to 2025 May 28 (548 days after first light). 
ATLAS $co$-band photometry was retrieved from the ATLAS forced photometry server\cite{Shingles2021TNSAN...7....1S}. Quality cuts\cite{Young2024zndo..10978968Y} and sigma clipping at the $5\sigma$ level were applied to filter outlier data points. A baseline was measured as the median flux in the time window well before ($>500$ days) discovery, and flux measurements were corrected for the baseline.
ZTF $gri$-band photometry was retrieved from the ZTF forced-photometry service\cite{Masci2019PASP..131a8003M,Masci2023arXiv230516279M} and processed with quality cuts\cite{Masci2023arXiv230516279M} and $5\sigma$ clipping. For each combination of the band, ZTF field, and CCD quadrant, flux measurements were corrected for the baseline, as similarly done for the ATLAS photometry. Flux uncertainties were corrected for the reduced $\chi^2$ distribution of PSF fits\cite{Masci2023arXiv230516279M}.

All the photometry in each band is weighted by uncertainty and stacked every 0.5 days. The detection significance ($\sigma$) of the ATLAS and ZTF forced photometry is determined from the ratio of measured flux ($F_\nu$) to its error ($F_{\nu,{\rm err}}$). For the measurements above and below $5\sigma$, we report their magnitudes ($-2.5\,{\rm log}_{10}F_\nu+{\rm ZP}$) and $5\sigma$ upper limits ($-2.5\,{\rm log}_{10}[5\times F_{\nu,{\rm err}}]+{\rm ZP}$), respectively, where ``ZP'' is the zero-point in the AB magnitude system. 
This yields a last nondetection at $21.6$ mag in the ZTF $r$ band on MJD 60,225.5 and first detection at $20.0$ mag in the ZTF $g$ band on MJD 60,228.4 (Fig.~\ref{fig:LC}). By taking the midpoint of the last nondetection and first detection, we estimate the epoch of first light to be MJD 60,227.0\,$\pm$\,1.4 (or 2023 October 10.0\,$\pm$\,1.4), with the uncertainty corresponding to the difference between the first light and detection. Even if we use the most conservative epoch of first light of the last nondetection, the difference is only $1.3$ rest-frame days, not affecting the main results of this paper. We find no significant detection ($\geq5\sigma$) before first light when stacking all the forced photometry every month. 


We fit a blackbody SED to every epoch of photometry containing at least two filters obtained within 2 days of each other to estimate the blackbody temperature and radius using \texttt{emcee}\cite{Foreman-Mackey2013PASP..125..306F}. 
The fitted blackbody SED is then integrated to obtain bolometric luminosity shown in Extended Data Fig.~\ref{EDfig:BB}. We note that the observed SED peaks are bluer than the optical coverage in the first $\sim30$ days ($\gtrsim8{,}000$\,K), and the fitted blackbody temperatures may be underestimated\cite{Valenti2016MNRAS.459.3939V,Arcavi2022ApJ...937...75} by up to $\sim5{,}000$\,K, which translates to a factor of $\sim2$ in the bolometric luminosity.
We further apply Gaussian Process (GP) regression\cite{Pedregosa2011JMLR...12.2825P} on the bolometric light curve, as shown in Extended Data Fig.~\ref{EDfig:BB}, with a combination of Mat\'{e}rn ($\nu=2.5$) and Constant kernels for model parameter constraints (``Light-curve modeling'' section). The peak luminosity ($L_p$) and recombination timescale ($t_f$) are measured from the maximum of the GP light curve and the minimum of its curvature, respectively, with associated $1\sigma$ uncertainties (Fig.~\ref{fig:mod}).

\subsection{Follow-up spectroscopy.}
\label{sec:spec}

We obtained fifteen optical spectra in total from 2024 March 3 to 2025 May 25 (134 to 545 days after first light): nine with the Binospec spectrograph\cite{Fabricant2019PASP..131g5004F} on the 6.5\,m MMT Observatory (Arizona, USA), two with the FLOYDS spectrograph on the LCO 2\,m FTN through GSP, two with the Kast double spectrograph\cite{miller1994kast} on the Shane 3\,m telescope at Lick Observatory (California, USA), and two with the Low Resolution Imaging Spectrometer (LRIS\cite{Oke1995PASP..107..375O}) on the 10\,m Keck I telescope at the W. M. Keck Observatory (Hawai`i, USA).
The Binospec observations utilized the 270 grating and $1''$-wide slit, providing a wavelength coverage of $\sim3{,}800-9{,}200$\,\AA\ with a resolution of $R = \lambda/\Delta \lambda \approx 1{,}500$. The Binospec spectra were extracted, reduced, and calibrated following standard procedures using PyRAF.
The FLOYDS observations utilized the $2''$ slit with a wavelength coverage of $\sim3{,}500-10{,}000$\,\AA\, and resolution of $R\approx400-700$. The FLOYDS spectra were reduced with the PyRAF-based \texttt{floyds\_pipeline}\cite{Valenti2014}.
The Kast observations utilized the $2''$ slit, D57 dichroic, 600/4310 grism, and 300/7500 grating, providing a combined wavelength coverage of $\sim3{,}500-10{,}500$\,\AA\, and resolution of $R\approx800$. The Kast spectra were reduced following standard techniques\cite{Silverman2012MNRAS.425.1789S} utilizing IRAF\cite{Tody1986SPIE..627..733T} routines and custom Python and IDL codes (\url{https://github.com/ishivvers/TheKastShiv}).
The LRIS observations utilized the $1''$ slit, D560 dichroic, 600/4000 grism, and 400/8500 grating, providing a combined wavelength coverage of $\sim3{,}200-10{,}200$\,\AA\ and resolution of $R\approx900$. The LRIS spectra were reduced with the IDL-based \texttt{LPipe}\cite{Perley2019PASP..131h4503P}  pipeline.
Additional flux calibration was applied to all the optical spectra using coeval photometry (Figs.~\ref{fig:LC}\,\&\,\ref{fig:spec}).


Additionally, we obtained a near-infrared (NIR) spectrum with the Gemini Near-InfraRed Spectrograph (GNIRS) on the 8.1\,m Gemini North telescope (Hawai`i, USA) on 2024 October 8 and 12 (335 and 339  days after first light). The observations utilized the cross-dispersed mode on the GNIRS short camera ($0.15''$\,pix$^{-1}$) with the 32\,line\,mm$^{-1}$ grating and $0.45''$ slit, providing a wavelength coverage of $\sim8{,}500-25{,}000$\,\AA\ with a resolution of $R\approx1{,}000$. Using the Python-based \texttt{PypeIt}\cite{2020JOSS....5.2308P} pipepline, the spectra were reduced for each night individually, flux calibrated with telluric standard stars observed on the same night, and then coadded with a telluric correction. Additional flux calibration was applied using the common wavelength range ($\sim8{,}500-9{,}200$\,\AA) with a coeval optical spectrum. The NIR spectrum is shown in Extended Data Fig.~\ref{EDfig:nir}.

We measure expansion velocities of H$\alpha$, H$\beta$, and Fe~{\sc ii} $\lambda$5,169 from the absorption minimum by fitting a P~Cygni profile (i.e., a positive and negative Gaussian, corresponding to the emission and absorption components, respectively) to each line in the spectra (Fig.~\ref{fig:spec}). We note an additional red emission component required to fit the line profiles during the light-curve tail phase ($\geq330$ d). Given the lack of narrow absorption features, we place an upper limit on the CSM velocity of $\lesssim170$\,km\,s$^{-1}$ from the highest Binospec resolution. We translate the difference between the observed minimum and the rest wavelength of the line to an expansion velocity with the Doppler formula. The line velocity evolution is shown in Extended Data Fig.~\ref{EDfig:vel}.
Using the residual flux from the SN line fits, we directly integrate the unresolved narrow host emission regions (i.e., H$\alpha$, H$\beta$, [O~{\sc ii}]\,$\lambda$3,727, [O~{\sc iii}]\,$\lambda$4,959, [O~{\sc iii}]\,$\lambda$5,007, [N~{\sc ii}]\,$\lambda$6,584, [S~{\sc ii}]\,$\lambda$6,717, and [S~{\sc ii}]\,$\lambda$6,731; Fig.~\ref{fig:class}) for the measurements of SN local environment (``Host galaxy'' section). We use the mean of the flux maximum wavelengths to measure a host redshift of $z = 0.0879$.

\subsection{Follow-up X-ray and radio observations.}
\label{sec:Xrad}

We retrieved the \textit{Neil G.~Gehrels Swift} X-Ray Telscope (XRT) observations of SN~2023vbw (ToO Target ID 16534; Proposer P. J. Pessi) taken on 2024 February 28 (130 days after first light) with a total on-source exposure time of 1,590\,s. We obtained a $3\sigma$ upper limit of $5.8\times10^{-3}$\,counts\,s$^{-1}$ ($0.3-10$\,keV) using the \textit{Swift}-XRT web tool\cite{Evans2007A&A...469..379E,Evans2009MNRAS.397.1177E}. With a Milky Way H~{\sc i} column density\cite{HI4PI2016A&A...594A.116H} of $3.81\times10^{20}$\,cm$^{-2}$ at the SN coordinates and assuming a power-law spectrum with a photon index of 2, the count rate is converted\cite{Mukai1993Legac...3...21M} to an unabsorbed flux limit of $F_{0.3-10\,{\rm keV}}< 2.2\times10^{-13}$\,erg\,s$^{-1}$\,cm$^{-2}$, corresponding to a luminosity limit of $L_{0.3-10\,{\rm keV}}<4.2\times10^{42}$\,erg\,s$^{-1}$.  


We performed a single K-band ($18-26$\,GHz) targeted follow-up observation using the NRAO Karl G. Jansky Very Large Array (VLA) (Program ID VLA/25A-460; PI W. W. Golay). The 1.5\,hr observation ($\sim0.83$\,hr on-source) was collected on 2025 April 19 (512 days after first light) when the VLA was in the most compact D-configuration. We reduced and analyzed the data with the VLA Calibration Pipeline (v2024.1.1.22) included in the NRAO Common Astronomy Software Applications (CASA, v6.6.1)\cite{CASATeam:2022}. We imaged the data with standard CASA {\tt tclean} parameters used for wideband imaging with maximum point-source detection sensitivity. No source was detected at the position of SN~2023vbw to a $3\sigma$ upper limit of $F_{22\,{\rm GHz}}<13$\,$\mu$Jy, corresponding to a luminosity limit of $L_{22\,{\rm GHz}} < 5.4\times10^{37}$\,erg\,s$^{-1}$. 

While the luminosity and temporal ranges probed by the XRT observations are not particularly constraining as compared to previous detections of SNe~II strongly interacting with CSM (e.g., $L_{0.3-8\,{\rm keV}}\approx10^{41}$\,erg\,s$^{-1}$ at $\gtrsim 500$ days after first light\cite{Chandra2018SSRv..214...27C}), our targeted VLA observations probe the ranges of the previous detections (e.g., $L_{8\,{\rm GHz}}\approx10^{37}$\,erg\,s$^{-1}$ at $\gtrsim 500$ days after first light\cite{Chandra2018SSRv..214...27C}). We discuss the VLA limit in the context of our CSM model (``Light-curve modeling'' section), and plan to obtain deeper ($\sim2\times$) observations in later epochs.


\subsection{Host galaxy.}
\label{sec:host}

We perform photometry on the host galaxy of SN~2023vbw using a customized version of \texttt{Blast}\cite{Blast}, called \texttt{FrankenBlast}\cite{FrankenBlast}. \texttt{FrankenBlast} collects all available images of the host in the \textit{Galaxy Evolution Explorer} (GALEX\cite{Galex}), the Panoramic Survey Telescope and Rapid Response System (Pan-STARRS\cite{PanSTARRS}), DESI Legacy Imaging Surveys, SDSS, \textit{Two-Micron All-Sky Survey} (2MASS\cite{2MASS}), and \textit{Wide-field Infrared Survey Explorer} (WISE\cite{WISE}). In each image, \texttt{FrankenBlast} constructs an elliptical aperture around the host using the \texttt{photutils} package\cite{photutils}. When there is a nondetection of the host (which is true for the GALEX FUV, SDSS $u$, all 2MASS filters, and WISE $w3$ and $w4$), \texttt{FrankenBlast} corrects the measured aperture size of the filter with the nearest wavelength by the full width at half-maximum (FWHM) intensity of the filter with the nondetection. In total, we collect photometry or upper limits for the host in 20 filters: GALEX FUV and NUV, SDSS $ugriz$, DESI $gz$, Pan-STARRS $grizy$, 2MASS $JHK$, and WISE $w1$--$w4$. 
In addition, we utilize an optical host spectrum taken with the long slit placed on both SN~2023vbw and the host nucleus during the MMT/Binospec observation on 2024 October 4. 


To model the stellar population properties of the host, we use stellar population modeling inference code \texttt{Prospector}\cite{Leja2019, Johnson2021ApJS..254...22J}, which, uniquely, can jointly fit photometry and spectroscopy of a galaxy. \texttt{Prospector} produces model SEDs from a given set of stellar population properties through \texttt{FSPS} and \texttt{python-FSPS}\cite{FSPS_2009, FSPS_2010}, which internally rely on \texttt{MIST} models and \texttt{MILES} spectral libraries\cite{MILES}. To determine posterior distributions on the stellar population properties of interest, we employ a nested sampling fitting routine, \texttt{dynesty}\cite{Dynesty}. Our \texttt{Prospector} model includes the Kroupa initial mass function (IMF)\cite{kroupaIMF}, the Kriek and Conroy dust emission model\cite{KriekandConroy13}, which measures an offset from the Calzetti dust attenuation curve\cite{calzetti2000} and the ratio of light attenuated from old to young stellar light, an infrared dust emission model\cite{DraineandLi07}, and a nebular emission model\cite{bdc+2017}. To probe realistic combinations of the stellar metallicity ($Z_*$) and total mass formed in the host, which is later converted to a stellar mass ($M_*$) using the star-formation history (SFH) of the host, we further apply a mass-metallicity relation\cite{gcb+05}. We model the observed spectral continuum with a 12th-order Chebyshev polynomial and apply a spectral smoothing model to normalize the continuum to the observed photometry. To fit the observed spectral line strengths, we determine a gas-phase metallicity ($Z_\textrm{gas}$) and apply a nebular marginalization template to marginalize over the observed emission lines. As we do not wish to overweight the spectrum in comparison to the photometry in the fit, we further employ a spectral noise inflation model and a pixel outlier model. Finally, we constrain the SFH of the host through a 7-bin nonparametric SFH model, which assumes a constant star-formation rate (SFR) in each age bin. The first two age bins range from 0 to 30 Myr and 30 to 100 Myr, and the final five are log-spaced to the age of the Universe at the host's redshift. We report the present-day SFR as the SFR determined from the first two age bins.


As shown in Extended Data Fig.~\ref{EDfig:SED}, we fit all photometric detections and upper limits along with the Binospec spectrum of the host at the redshift of $0.0879$ (``Follow-up spectroscopy'' section). We find that the host has $\log(M_*/{\rm M}_\odot)=9.28^{+0.05}_{-0.05}$, $\log(Z_*/{\rm Z}_\odot)=-1.09^{+0.14}_{-0.11}$, and $\log(Z_\textrm{gas}/{\rm Z}_\odot)=-0.65^{+0.08}_{-0.05}$. We measure a dust extinction of $A_V=0.09^{+0.13}_{-0.07}$~mag from the total optical depth of dust attenuated from old and young stellar light. We find that the SFH is declining, with the most recent star-formation burst in the host $\sim 2-3$~Gyr before SN 2023vbw. Using the SFH, we calculate a mass-weighted stellar population age of $5.55^{+0.91}_{-1.0}$~Gyr. Finally, we determine a present-day ${\rm SFR}=0.19^{+0.15}_{-0.06}\,{\rm M}_\odot$~yr$^{-1}$ and specific SFR, ${\rm log(sSFR)}=-9.69^{+0.13}_{-0.18}$~yr$^{-1}$. Despite the declining SFH, the host is still considered actively star-forming, given its redshift and sSFR\cite{tacchella2022}.


In addition, we make measurements of the SN local environment with the host emission lines in the SN spectra (Fig~\ref{fig:class}; ``Follow-up spectroscopy'' section). After correcting for reddening using the Balmer decrement ($E(B-V)=0.27\pm0.04$ mag), we find a SFR of $0.07\pm0.02\,{\rm M}_\odot\,{\rm yr}^{-1}$ from the H$\alpha$ line luminosity\cite{Kennicutt1998ARA&A..36..189K} and a gas-phase metallicity of $12+{\rm log(O/H)}=8.21\pm0.02$ (or $\approx0.3\,{\rm Z}_\odot$ assuming a solar metallically\cite{Asplund2021A&A...653A.141A} of $12+{\rm log(O/H)}=8.69\pm0.04$) from the flux ratios\cite{Marino2013A&A...559A.114M,Bianco2016A&C....16...54B} of [N~{\sc ii}]\,$\lambda$6,584 / H$\beta$ and ([O~{\sc iii}]\,$\lambda$4,959 + [O~{\sc iii}]\,$\lambda$5,007) / H$\beta$. Together with the global host measurements, the SN local measurements are well within the expectations for PISNe\cite{Briel2024MNRAS.533.3907B}.

\subsection{Light-curve modeling.}
\label{sec:mod}
With the lack of comprehensive numerical light-curve models spanning PISN parameters (see the discussion at the end of this section), we compare the bolometric light curve of SN 2023vbw to a parameter grid obtained by a semi-analytical model constructed for SNe~II\cite{Pumo2025MNRAS.538..223P} to infer the properties of its progenitor and explosion. The model assumes homologously expanding SN ejecta with uniform density (also supported by multi-dimensional simulations of hydrogen-rich PISN models\cite{Chen2020ApJ...897..152C}), and takes into account the time-dependent hydrogen recombination in the ejecta as well as differences in opacity within the ejecta due to recombination. The model can also incorporate heating sources in the ejecta, such as radioactive decay of $^{56}$Ni (and its decay product $^{56}$Co) synthesized in the explosion. 
The morphology of the main peak of SN~1987A-like SNe (Fig.~\ref{fig:LC}), with a slow rise and steep decline, is governed by an interplay of both recombination and heating. For accurately constraining the ejecta mass, the current model is favoured over other widely used semi-analytical modeling tools\cite{MOSFiT,REDBACK}, which assume uniform opacity throughout the ejecta for simplicity.

The main parameters of the model are the ejecta mass ($M_{\rm ej}$), kinetic energy ($E_{\rm ej}$), progenitor radius ($R_0$), and $^{56}$Ni mass ($M_{\rm Ni}$) in the ejecta. $M_{\rm ej}$ and $E_{\rm ej}$ are related by the outermost scale velocity of the ejecta ($v_{\rm sc}$; the fastest ejecta velocity in homologous phase), as $E_{\rm ej}=0.3M_{\rm ej}v_{\rm sc}^2$, and we adopt $v_{\rm sc}=8{,}000$\,km\,s$^{-1}$ based on the H$\alpha$ P Cygni line and the earliest blackbody radius evolution that tracks the fastest part of the SN ejecta (Extended Data Figs.~\ref{EDfig:BB}\,\&\,\ref{EDfig:vel}). For the distribution of $^{56}$Ni in the ejecta, we choose the model's ``EXP+SOE" prescription that well reproduces the light-curve morphology of SN 1987A\cite{Pumo2025MNRAS.538..223P}. This distribution sets the $^{56}$Ni to be within the central $\sim 10\%$ in mass, which is also in reasonable agreement with nucleosynthesis predictions of numerical PISN models\cite{Joggerst2011ApJ...728..129J,Dessart2013MNRAS.428.3227D,Chen2014ApJ...792...44,Gilmer2017ApJ...846..100G,Chen2020ApJ...897..152C}.

We constrain the parameters that reproduce two characteristic quantities in the light-curve main peak: the peak luminosity ($L_p$) and the end of the optically thick phase ($t_f$). Fig. \ref{fig:mod} shows the constraint on $M_{\rm ej}$ and $M_{\rm Ni}$ for a radius $R_0=100~{\rm R}_\odot$, typical for BSG progenitors. For such compact progenitors, radioactive decay of $^{56}$Ni is mainly responsible for powering the main peak, and thus $L_p$ mostly constrains $M_{\rm Ni}$. On the other hand, $t_f$ is governed by both $M_{\rm ej}$ that sets photon diffusion and $M_{\rm Ni}$ that sets the heating of the ejecta, creating a diagonal contour. These two constraints yield values of $M_{\rm ej}$ and $M_{\rm Ni}$ that agree with theoretical expectations of PISNe\cite{Heger2002ApJ...567..532H,Heger2003ApJ...591..288H,Takahashi2018ApJ...857..111T,Kasen2011ApJ...734..102K,Dessart2013MNRAS.428.3227D}.

As $R_0$ increases, the SN ejecta are subject to less adiabatic loss that reduces their internal energy as $E_{\rm int}\propto (R_0/v_{\rm sc}t)$. Thus, a higher $R_0$ leads to an increasing contribution from the internal energy generated by the explosion\cite{Popov1993,Kasen2009}. As shown in Extended Data Fig.~\ref{EDfig:mod_var} with two variations of $R_0$ ($50$ and $500~{\rm R}_\odot$), while reducing $R_0$ makes little change from our fiducial assumption ($100~{\rm R}_\odot$), increasing $R_0$ starts to raise the luminosity at early times, flattening the rise, an effect also seen for lower-mass explosions\cite{Taddia2016A&A...588A...5T}. For much larger radii ($\gtrsim 1{,}000\, {\rm R}_\odot$) expected for RSGs, theoretical models\cite{Kasen2011ApJ...734..102K,Dessart2013MNRAS.428.3227D} predict a nearly plateau-like light-curve morphology, analogous to SNe~IIP, but contrary to SN~2023vbw's slow rise. Hence, we favour a BSG over an RSG progenitor. 

The early phase of the light curve, as well as the tail after the rapid decline, show an excess compared to the model light curve powered solely by radioactive decay (Fig.~\ref{fig:mod}). Furthermore, multicomponent hydrogen lines emerge at late times (Fig.~\ref{fig:spec} \& Extended Data Fig.~\ref{EDfig:nir}), while such signatures are absent during the main peak. These features could be explained by shock interaction of the SN ejecta and a pre-existing disc-like CSM\cite{Kurf2020A&A...642A.214K}, where CSM interaction is hidden by the optically thick SN ejecta during the main peak\cite{Smith2015MNRAS.449.1876S}. Both the early ($\lesssim 40$ d) and tail ($\gtrsim 300$ d) emission can be fit by a single power-law component $L_{\rm CSM}\approx 4.5\times 10^{42}\ {\rm\,erg\,s^{-1}}(t/100\ {\rm day})^{-3/8}$, whose power-law index is expected\cite{Moriya2013MNRAS.435.1520M} for BSG ejecta colliding with a CSM with a wind-like density profile $\rho(r)=\dot{M}_{\rm CSM}/(4\pi r^2f_\Omega v_{\rm CSM})$, where $f_\Omega$ is the covering fraction of the CSM. The collision forms a shock that propagates the CSM with a velocity $v_{\rm sh}\propto t^{-1/8}$ for BSG ejecta, which is nearly constant in time. The luminosity generated by the shock interaction can be expressed as\cite{Moriya2013MNRAS.435.1520M}
\begin{equation}
    L_{\rm CSM} = \frac{\epsilon_{\rm rad}}{2}(4\pi r^2 f_\Omega)\rho v_{\rm sh}^3 = \frac{\epsilon_{\rm rad}}{2}\frac{\dot{M}_{\rm CSM}}{v_{\rm CSM}}v_{\rm sh}^3,
\end{equation}
where $\epsilon_{\rm rad}$ is the conversion efficiency to radiation, which is expected to be inefficient with $\epsilon_{\rm rad}\approx 0.1$ due to adiabatic losses as a large fraction of photons generated by the interaction are trapped in the optically thick SN ejecta\cite{Khatami2024ApJ...972..140K}. Adopting a shock velocity of $v_{\rm sh} \approx 8{,}000$\,km\,s$^{-1}$ based on the blackbody radius evolution up to 40 d (Extended Data Fig.~\ref{EDfig:BB}), this gives a rough estimate of
\begin{equation}
    \dot{M}_{\rm CSM} \approx 0.05~M_\odot\ {\rm yr}^{-1}\left(\frac{\epsilon_{\rm rad}}{0.1}\right)^{-1}\left(\frac{v_{\rm CSM}}{170\ {\rm km\ s^{-1}}}\right)\left(\frac{v_{\rm sh}}{8{,}000\ {\rm km\ s^{-1}}}\right)^{-3}.
\end{equation}
Note that the same CSM underproduces the luminosity in the main peak (Fig. \ref{fig:mod}), and our interpretation of a large $M_{\rm Ni}$ for the main peak is unaffected to within a few 10\% (``Alternative scenario" section). 

The late-time hydrogen lines with a wavelength shift of $\sim 3{,}000$\,km\,s$^{-1}$ (Fig.~\ref{fig:spec}) imply ongoing shock interaction with the disc-like CSM\cite{Smith2015MNRAS.449.1876S,Andrews2018MNRAS.477...74A,Sollerman2019A&A...621A..30S,Kurf2020A&A...642A.214K}. While the strong P Cygni absorption from the SN ejecta precludes robustly constraining the bluer component of CSM interaction, a prominent redshifted component indicates strong interaction at the farther side of the disc from the observer. From viewing-angle effects, the velocity shift could be interpreted as a lower limit on the shock velocity. Extrapolating the shock velocity evolution of $v_{\rm sh}\propto t^{-1/8}$ to $1$\,yr from explosion gives an estimate of $v_{\rm sh}\approx 6{,}000$\,km\,s$^{-1}$, consistent with the above lower limit. This implies the observer to be neither face-on nor edge-on, but at an intermediate angle with respect to the disc.

From the $\dot{M}_{\rm CSM}/v_{\rm CSM}$ inferred above, we can estimate the free-free attenuation of radio waves by the unshocked CSM\cite{Murase2014MNRAS.440.2528}. For spherically symmetric CSM located at $r\geq v_{\rm sh}t$ with characteristic electron temperature of $T_e\approx 10^4$ K, this yields a free-free optical depth at 22 GHz of $\tau_{\rm 22 GHz}\approx 100(T_e/10^4\ {\rm K})^{-1.5}(v_{\rm sh}/8{,}000\ {\rm km\ s^{-1}})^{-3}(t/512\ {\rm day})^{-3}$, consistent with the VLA nondetection at 512 days (``Follow-up X-ray and radio'' section). For a disc-like CSM configuration, a face-on observer could have a reduced $\tau$, while an edge-on observer could have an enhanced $\tau$.

Finally, we check the robustness of our semi-analytical parameter estimation against existing numerical PISN models. In Extended Data Fig.~\ref{EDfig:PISN_mod}, we compare the light curve and line velocity evolution of SN~2023vbw with the nearest neighbor PISN models (Fig.~\ref{fig:mod}), namely the BSG and RSG models: B200 and R175 from K11\cite{Kasen2011ApJ...734..102K} and B190 and R190 from D13\cite{Dessart2013MNRAS.428.3227D}, with available light curves and photospheric velocities. Since the PISN models assume no CSM interaction despite their very massive progenitors, they are fainter than SN~2023vbw in late tail phase. For fair comparison, we also show the model light curves with the addition of the same CSM contribution estimated for SN~2023vbw. Due to large $R_0$ of $2{,}500$ and $4{,}000\,{\rm R}_\odot$, the RSG models show luminous and extended shock-cooling phase, which is inconsistent with the last nondetection. On the other hand, the BSG models show the light-curve morphology that is consistent with SN~2023vbw, albeit with longer durations. The BSG models also better match the Fe~{\sc ii}\,$\lambda 5{,}169$ velocity (i.e., proxy for the photospheric velocity; Extended Data Fig.~\ref{EDfig:BB}) evolution than the RSG models, albeit with slower velocities. As the velocity scales with $(E_{\rm ej}/M_{\rm ej})^{1/2}$, the lower $E_{\rm ej}/M_{\rm ej}$ of the BSG models ($\sim0.14$ and $0.26\,{\rm B \, M}_\odot^{-1}$ for B200 and B190, respectively, where ${\rm B}=10^{51}\,{\rm erg}$) than the semi-analytical estimate of SN~2023vbw ($= 0.3 v_{\rm sc}^2 \approx 0.38\,{\rm B \, M}_\odot^{-1}$) result in their slower velocities, and consequently longer light-curve durations due to photon diffusion. In summary, the overall light curve and velocity evolution of SN~2023vbw are reasonably well reproduced by the numerical PISN models from BSG progenitors with the addition of CSM interaction, where the shorter light-curve duration and faster velocity of SN~2023vbw are attributed to its higher $E_{\rm ej}/M_{\rm ej}$.


\subsection{Possible formation channels of the progenitor and CSM.}
\label{sec:form}

The modeling of SN 2023vbw suggests a very massive ($\gtrsim 100\,{\rm M}_\odot$) and compact (BSG-like) progenitor, embedded in a continuous disc-like CSM with orders of magnitude higher density than radiatively-driven winds. Such conditions are analogous to SN 1987A\cite{1991Natur.350..683C}, an explosion of a BSG within a detached, equatorial ring at a radius of $\approx 0.2$ pc and expansion velocity of $10$ km s$^{-1}$. The BSG progenitor and the ring have both been explained by a merger occurring $(0.2\ {\rm pc}/10\ {\rm km\ s^{-1}})\sim 20$ kyr before core-collapse, with the ring created by equatorial mass loss from the rotating merger product as its envelope thermally contracts to a BSG\cite{Morris2007Sci...315.1103M,2009MNRAS.399..515M,Menon2017MNRAS.469.4649M}. The upper limit on the CSM velocity ($\lesssim170$\,km\,s$^{-1}$) of SN 2023vbw in the post-peak interacting phase is consistent with such a slow, equatorial outflow.

We can observationally constrain the timing of the mass ejection for SN 2023vbw, agnostic of its detailed origin. The persistent interaction signatures up to 550 days after the SN gives a lower limit on the extent of the CSM as
\begin{equation}
R_{\rm CSM}\gtrsim v_{\rm sh}\times 550\ {\rm day}\approx 4\times 10^{16}\ {\rm cm}.    
\end{equation}
A loose upper limit of $R_{\rm CSM}$ can be obtained by restricting the total CSM mass to be less than the total hydrogen mass available in the envelope $M_{\rm H, env}$, 
\begin{equation}
    R_{\rm CSM}\lesssim \frac{M_{\rm H, env}}{\dot{M}_{\rm CSM}}v_{\rm CSM} \sim 9\times 10^{17}\ {\rm cm}\left(\frac{M_{\rm H, env}}{100\,{\rm M_\odot}}\right) \left(\frac{\epsilon_{\rm rad}}{0.1}\right)\left(\frac{v_{\rm sh}}{8{,}000\ {\rm km\ s^{-1}}}\right)^{3},
\end{equation}
noting the dependence on $v_{\rm CSM}$ canceled out. The onset of the mass ejection is hence constrained to be $\sim 1-30$ kyr $(v_{\rm CSM}/10\ {\rm km\ s^{-1}})^{-1}$ before the SN, or $0.1-30$ kyr for a plausible range of $v_{\rm CSM}\approx 10-100\ {\rm km\ s^{-1}}$. This is around the end of core helium burning stage for massive stars, when the star expands as it transitions to the helium shell burning stage.

In the following, we suggest that the post-core helium burning (Case C) merger channel, widely accepted for the progenitor of SN 1987A, could explain the observed CSM properties around the BSG progenitor of SN 2023vbw. As the merger product contracts to become a BSG, the outermost envelope rotating beyond the critical specific angular momentum (which scales with the stellar radius as $R_*^{1/2}$) will be lost by centrifugal forces as an equatorial wind\cite{Morris2007Sci...315.1103M}. If the star is not disrupted, this mass loss would continue for a few thermal (Kelvin-Helmholtz) timescales $t_{\rm KH}$, until the star radiates its excess thermal energy from the merger and settles its radius. The thermal timescale $t_{\rm KH}$ is
\begin{equation}
 t_{\rm KH} \approx \frac{GM_*^2}{R_*L_*} \sim 2\ {\rm kyr}\left(\frac{M_*}{200\,{\rm M}_\odot}\right)\left(\frac{R_*}{100\,{\rm R}_\odot}\right)^{-1}\left(\frac{L_*}{L_{\rm Edd}}\right)^{-1},
\end{equation}
where $G$ is the gravitational constant, $M_*$ is the stellar mass, and $L_*$ is the stellar luminosity in which we scale by the Eddington limit $L_{\rm Edd}\approx 3\times 10^{40}\ {\rm erg\ s^{-1}}(M_*/200\,{\rm M}_\odot)$ as $L_*\approx L_{\rm Edd}$ for very massive stars. We note that $t_{\rm KH}$ becomes larger as the star contracts, and the mass loss phase is thus spent longest during the BSG phase.

The duration of the thermal contraction phase (a few $t_{\rm KH}$) can be compared with the remaining lifetime of the progenitor, with an important point being the much shorter remaining lifetime from helium shell burning of PISN progenitors ($\sim$\, kyr\cite{Dessart2013MNRAS.428.3227D,Takahashi2018ApJ...857..111T,Takahashi2018ApJ...863..153T}) compared to typical massive stars like the progenitor of SN 1987A ($\gtrsim 10$\,kyr). For the case of SN 1987A, the thermal contraction phase is much shorter than the remaining lifetime, such that the equatorial CSM becomes detached from the star at core-collapse, forming the ring we observe today. On the other hand, a PISN progenitor would explode during the thermal contraction phase due to its shorter lifetime, which leads to an equatorial CSM that exists from just outside the star to distances traversed by $\sim$\, kyr, consistent with the CSM constraints of SN 2023vbw. Stated another way, for the Case C merger channel, the CSM existing very close to the star requires a very short lifetime of $\lesssim$ a few $t_{\rm KH}$ from merger to explosion, achievable only for very massive stars like PISN progenitors.

While we suggest a PISN progenitor that underwent Case C merger as the most plausible possibility of the progenitor and CSM of SN 2023vbw, uncertainties remain in the evolution of very massive stars potentially leading to PISNe, such as the red/blue supergiant dichotomy\cite{2021MNRAS.504..146V,2024MNRAS.529.2980W} and the timing of potential binary mergers\cite{2008AIPC..990..230D,2025RNAAS...9...75B}. We encourage future stellar/binary evolution simulations for a viable model of SN 2023vbw that simultaneously explains the bright, long-rising light curve and the nearby, dense CSM at distance of $\sim 10^{16}$ cm from the progenitor.

\subsection{Alternative scenarios.}
\label{sec:alt}

In this paper, we have considered the PISN scenario of the entire star exploding and synthesizing large amounts of $^{56}$Ni, where the powering source of the main peak is the radioactive decay of $^{56}$Ni and $^{56}$Co. In the literature, however, alternative mechanisms to luminous SNe have been proposed, where an iron CCSN is heated by energy injection from a central compact object, either a rapidly spinning neutron star (NS)\cite{Kasen2010ApJ...717..245K} or an accreting compact object (NS or black hole)\cite{Dexter2013}.


Regardless of the nature of the heating mechanism, the rise of the light curve is governed by photon diffusion, which directly relates to $M_{\rm ej}$. Hence, we expect the light curve to still require a large $M_{\rm ej}$ ($\gtrsim 100~{\rm M}_\odot$), and thus a similarly massive progenitor. As the time dependence of fallback accretion is uncertain, we consider the case of energy injection by NS dipole radiation where the heating rate in the ejecta is given as $L_{\rm sd} = L_0[1+(t/t_{\rm sd})]^{-2}$, with $t_{\rm sd}$ being the spindown timescale and $L_0$ being the initial spindown luminosity. Here, we neglect heating due to $^{56}$Ni decay, as an iron CCSN from a canonical neutrino-powered explosion\cite{Sukhbold2016ApJ...821...38S} would produce only $\lesssim 0.1~{\rm M}_\odot$ of $^{56}$Ni (Fig.~\ref{fig:mod}), much less than the $\sim 1~{\rm M}_\odot$ required to dominantly power the light curve of SN~2023vbw by $^{56}$Ni decay. Our modeling similarly gives the constraint on the ejecta mass of $M_{\rm ej}\approx 100-300~{\rm M}_\odot$, with a best fit of $\sim 200~{\rm M}_\odot$. The best-fit values of the spindown parameters are $L_0\approx 4\times 10^{43}$\,erg\,s$^{-1}$ and $t_{\rm sd}\approx 150$ days, corresponding to an initial spin period of $\sim 8$\,ms and surface magnetic field of $\sim 10^{14}$\,G for an NS mass of $M_{\rm NS}=1.4~{\rm M}_\odot$ and radius of $12$\,km.

A rapidly spinning NS powering the SN can be ruled out with high confidence based on the inferred $M_{\rm ej}$. An NS remnant from a massive progenitor of mass $M_*=M_{\rm ej}+M_{\rm NS}\gtrsim 100~{\rm M}_\odot$ is difficult to reconcile with stellar evolution theory\cite{Heger2003ApJ...591..288H,Takahashi2018ApJ...857..111T,Takahashi2018ApJ...863..153T}, that instead predicts either a black hole remnant or a PISN with no remnant (Fig.~\ref{fig:mod}). Although binary evolution with mass transfer during main sequence (Case A) may allow NS formation from such a massive progenitor due to hydrogen-rich envelope stripping and helium core reduction\cite{Belczynski2008ApJ...685..400B}, the resultant explosion would be a hydrogen-poor supernova unlike SN~2023vbw. The inferred mass further indicates $E_{\rm ej}\approx 10^{53}$\,erg, two orders of magnitude larger than both the explosion energy of canonical iron CCSNe and the rotational energy of an NS with spin period of 8\,ms. Finally, a hydrogen-rich progenitor is expected to have efficient angular momentum transport from the core to the envelope that rapidly spins down the core, preventing the formation of a fast-rotating NS upon core-collapse\cite{Ma2019MNRAS.488.4338}.

The possibility of fallback accretion is difficult to rule out from stellar evolution theory, given uncertainties regarding how accretion onto the remnant compact object proceeds and how efficiently energy can be extracted from the accretion disc via an outflow or jet. The late-time light curve can also be reasonably fit by a $t^{-5/3}$ decay, a scaling expected for fallback accretion at late times. However, the large $E_{\rm ej}\approx 10^{53}$\,erg, if supplied from fallback accretion, requires an extreme fallback of $E_{\rm ej}/(\epsilon_{\rm inj} c^2)\approx 50~{\rm M}_\odot(\epsilon_{\rm inj}/10^{-3})^{-1}$, where $c$ is the speed of light and $\epsilon_{\rm inj}$ is the energy injection efficiency from accretion taken to be $10^{-3}$ here\cite{Dexter2013}. 

Even if such a huge amount of fallback is possible, sustained heating by accretion for $\gtrsim 100$ days, as required from the spectral evolution of SN~2023vbw (Fig.~\ref{fig:spec}), is implausible on energetic grounds. For a mass accretion rate of $\dot{M}_{\rm fb}(t)=\dot{M}_0(t/t_0)^{-5/3}$, the remaining mass bound to the remnant at time $t$ is $M_{\rm fb}=\int_t^{\infty} \dot{M}_{\rm fb}(t')\, dt' =(3\dot{M}_0t_0/2)(t/t_0)^{-2/3}$, and a characteristic radius set by the marginally bound material is $R_{\rm fb}\approx (GM_\bullet t^2)^{1/3}$, where $G$ is the gravitational constant and $M_\bullet$ is the remnant mass. Accretion onto the central remnant will deposit the bound material with an energy of $E_{\rm dep}\approx f(\epsilon_{\rm inj} \dot{M}_{\rm fb}c^2) (R_{\rm fb}/v_{\rm h})$, where $v_{\rm h}\approx (\epsilon_{\rm inj}\dot{M}_{\rm fb}c^2 R_{\rm fb}/3M_{\rm fb}v_{\rm out})^{1/2}$ is the velocity of the head of the outflow propagating the fallback material, and $f\gtrsim 0.03$ is the efficiency of thermalizing the outflow that could be highly magnetized\cite{Dexter2013,Quataert2012MNRAS.419L...1}. The ratio of $E_{\rm dep}$ to the binding energy $E_{\rm bind}\approx GM_\bullet M_{\rm fb}/R_{\rm fb}$ is
\begin{equation}
    \frac{E_{\rm dep}}{E_{\rm bind}}\approx f\sqrt{\frac{2\epsilon_{\rm inj} v_{\rm out}c^2 t}{GM_\bullet}} \sim 60\left(\frac{f}{0.03}\right)\left(\frac{\epsilon_{\rm inj}}{10^{-3}}\right)^{1/2}\left(\frac{v_{\rm out}}{0.1c}\right)^{1/2}\left(\frac{M_\bullet}{100~{\rm M}_\odot}\right)^{-1/2}\left(\frac{t}{100\ {\rm day}}\right)^{1/2}\, .
\end{equation}
Thus, the energy deposition by accretion is well in excess of the binding energy much before 100 days, effectively shutting off further accretion. Although a much smaller $\epsilon_{\rm inj}$ could make the accretion last longer, it would exacerbate the problem of requiring an even larger fallback mass to explain the large $E_{\rm ej}$. A highly collimated energy injection with an opening angle of $\theta\ll 1^\circ$ increases $v_h$ as $v_h\propto \theta^{-1}$ and reduces $E_{\rm dep}$ as $E_{\rm dep}\propto \theta$, potentially making continued accretion possible; however, whether such a strongly collimated outflow of $\theta\ll 1^\circ$ can be launched is highly uncertain.

Finally, we verify that the specific CSM profile invoked to match the early and late emission cannot power the main peak of SN~2023vbw. Regardless of the detailed CSM density profile, powering the radiated energy of $E_{\rm rad}\approx 3\times 10^{50}$\,erg in a regular iron CCSN interacting with disc-like CSM is energetically difficult to achieve. Such disc-like CSM likely originates from a centrifugally driven wind or binary interaction/merger (``Possible formation channels of the progenitor and CSM" section), and we expect the covering fraction of the disc to be set by hydrostatic balance as $f_\Omega\approx c_s/v_{\rm orb}$, where $c_s$ and $v_{\rm orb}$ are the isothermal sound speed and orbital velocity of the disc material\cite{Shu1979ApJ...229..223,Pejcha2016MNRAS.455.4351}, respectively. As $c_s$ is set by the stellar surface temperature ($T_*\approx (1-2)\times 10^4$ K for a BSG, i.e., $c_s\approx 10$\,km\,s$^{-1}\sqrt{T_*/10^4\ {\rm K}}$), and $v_{\rm orb}$ is comparable to the progenitor's surface escape speed $v_{\rm esc}=\sqrt{GM_*/R_0}\approx 140\ {\rm km\ s^{-1}}(M_*/10~{\rm M}_\odot)^{1/2}(R_0/100~{\rm R}_\odot)^{-1/2}$, we expect $f_\Omega \lesssim 10\%$ for $M_*=10-100~{\rm M}_\odot$. Thus, even for the most optimistic $\epsilon_{\rm rad}$ of unity, we require an explosion energy of $E_{\rm ej}=E_{\rm rad}/f_\Omega\approx(3-10)\times 10^{51}$\,erg, which is higher than predictions\cite{Sukhbold2016ApJ...821...38S} for iron CCSNe, of $\sim(0.1-2)\times 10^{51}$ erg. The actual $\epsilon_{\rm rad}$ is expected to be much lower since the interaction region is well embedded below the photosphere (as supported by the broad P~Cygni profiles without narrow interaction components; Fig.~\ref{fig:spec}), resulting in significant adiabatic losses due to reprocessing by the SN ejecta\cite{Khatami2024ApJ...972..140K}, which will make the required $E_{\rm ej}$ even higher. Moreover, in such a case, CSM interaction virtually acts as a central heating source where the light-curve peak is governed by photon diffusion, as similarly discussed in the scenarios with compact objects, requiring $M_{\rm ej}\gtrsim 100~{\rm M}_\odot$ and $E_{\rm ej}\approx 10^{53}$ erg, which again strongly disfavors a normal iron CCSN.

We therefore conclude that all possible alternative scenarios are not viable to explain the observed properties of SN~2023vbw, leaving the pair-instability origin the most natural one.

\end{methods}

%




 
\begin{addendum}

\item[Data Availability] The data that support the plots within this paper and other findings of this study will be available from the corresponding author upon reasonable request.


\item[Code Availability] The light-curve modeling code will be available upon publication.


\end{addendum}


\clearpage

\begin{addendum}

\item[Extended Data]

\begin{EDfigure}
\centering
\includegraphics[width=0.99\textwidth]{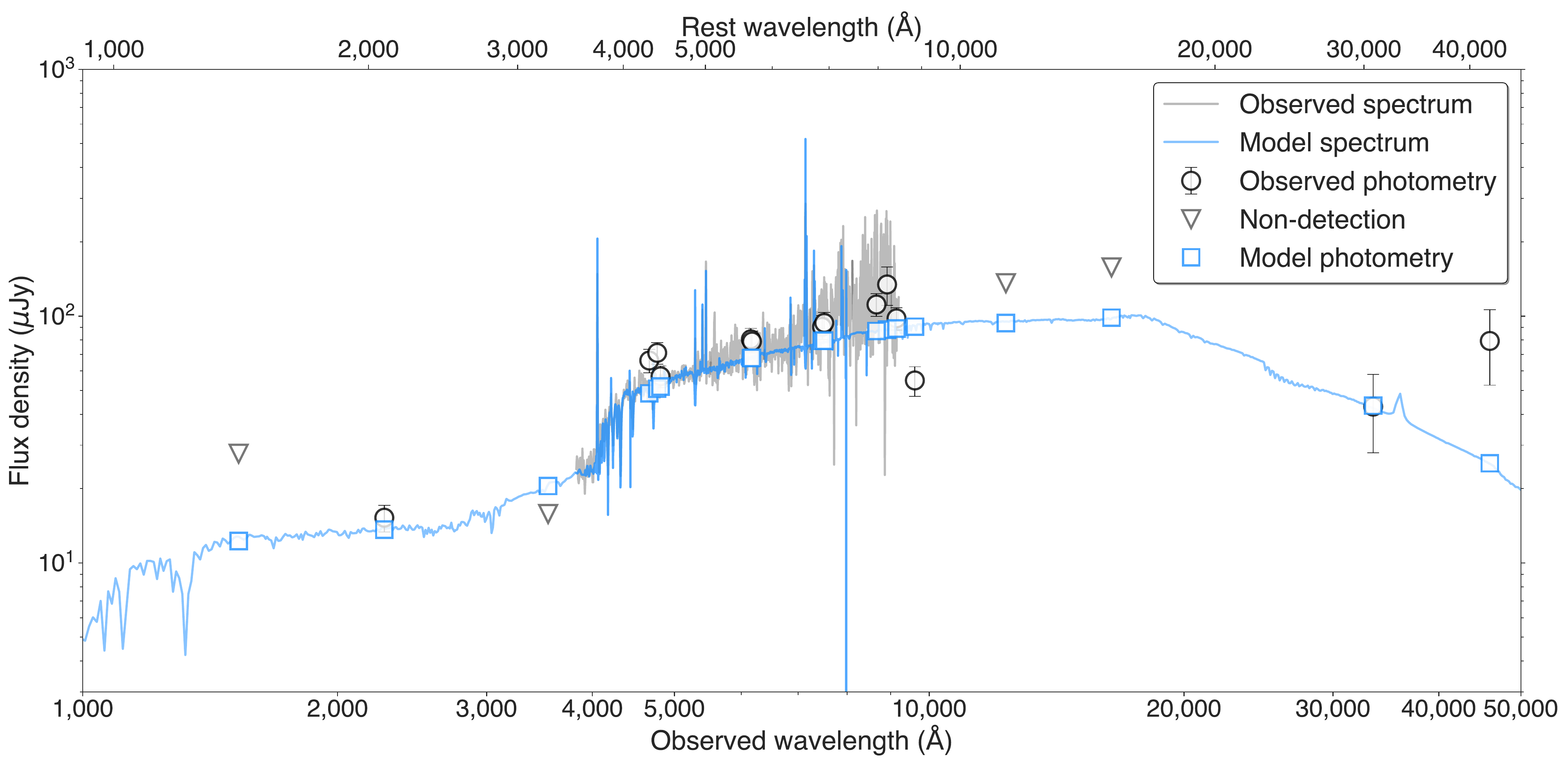}
\caption{
\textbf{Host-galaxy SED and model fit.}
The observed GALEX, Pan-STARRS, DESI, SDSS, 2MASS, WISE photometry, and MMT/Binospec spectrum of the host galaxy along with the \texttt{Prospector} model fit. Error bars denote $1\sigma$ uncertainties. The model-inferred parameters are listed in Methods
(``Host galaxy'' section).
\label{EDfig:SED}
}
\end{EDfigure}

\newpage

\begin{EDfigure}
\centering
\includegraphics[width=0.63\textwidth]{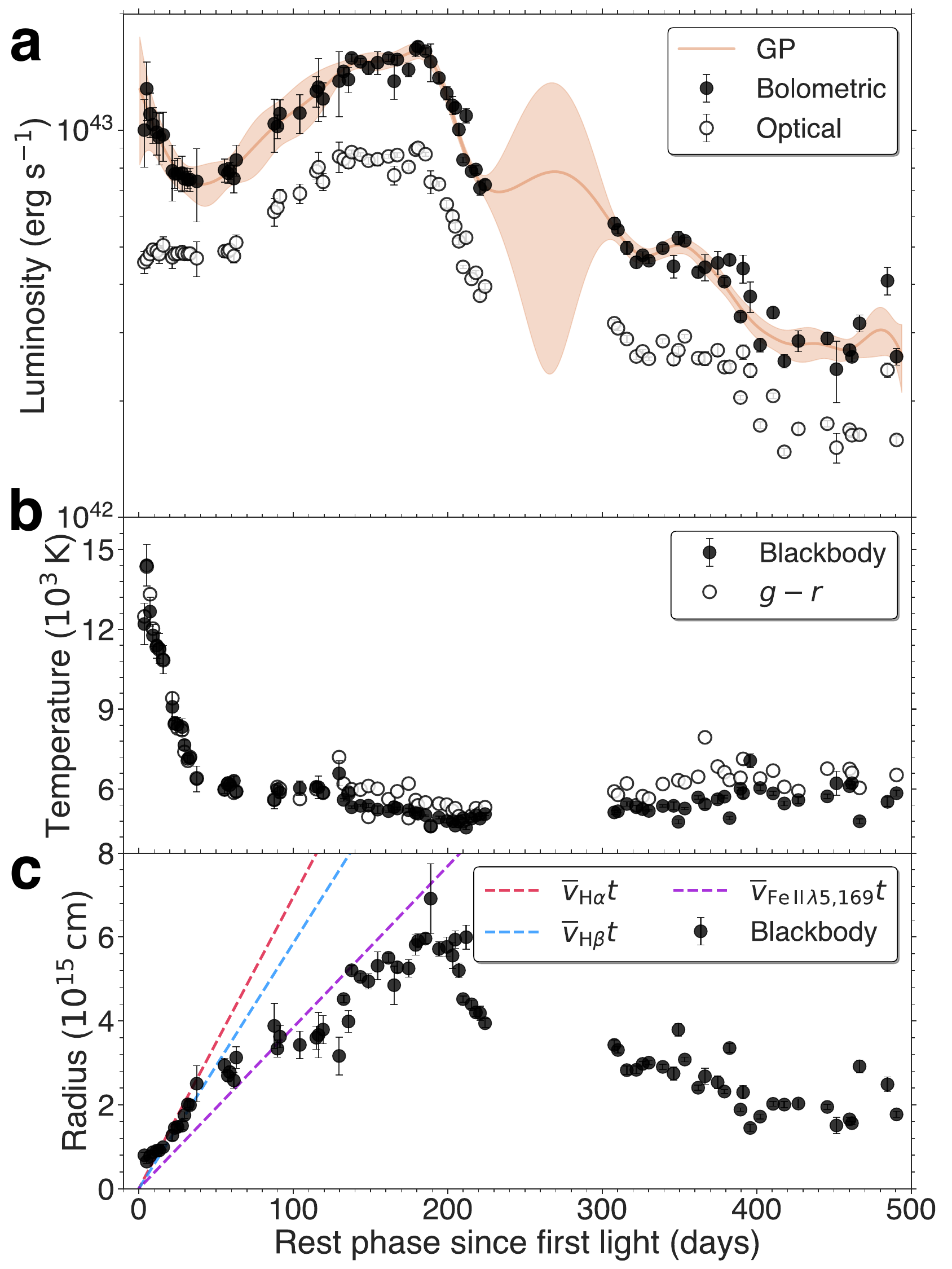}
\caption{
\textbf{Luminosity and blackbody evolution.}
\textbf{a,} The bolometric and optical ($3{,}250-8{,}900$\,\AA) luminosity along with the GP regression fit (``Follow-up photometry'' section). Error bars and shaded region denote $1\sigma$ uncertainties.
\textbf{b, c,} The blackbody temperature and radius evolution. Also shown are the $g-r$ colour temperature (Fig.~\ref{fig:LC}) and homologous expansion with the mean H$\alpha$, H$\beta$, and Fe~{\sc ii}\,$\lambda 5{,}169$ velocities during the light-curve rise (Extended Data Fig.~\ref{EDfig:vel}). The early rapid radius expansion has a similar velocity to H$\alpha$, while the following gradual expansion has a similar velocity to Fe~{\sc ii}\,$\lambda 5{,}169$. The radius then recedes after the light-curve peak.
\label{EDfig:BB}
}
\end{EDfigure}

\newpage

\begin{EDfigure}
\centering
\includegraphics[width=0.99\textwidth]{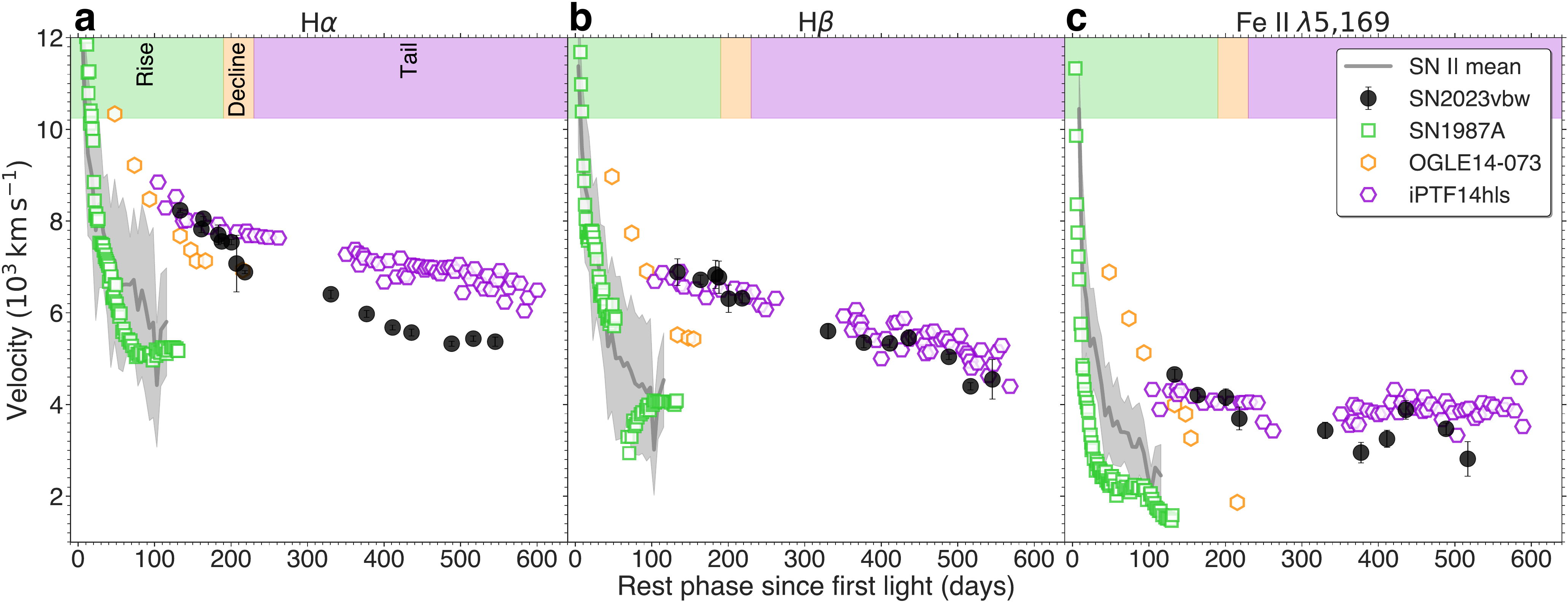}
\caption{
\textbf{Line velocity evolution.} 
\textbf{a--c,} Comparison of the H$\alpha$, H$\beta$, and Fe~{\sc ii}\,$\lambda 5{,}169$ velocity evolution of SN~2023vbw (``Follow-up spectroscopy'' section) with the mean SN~II velocity\cite{Guti2017ApJ...850...89G}, SN~1987A\cite{Phillips1988AJ.....95.1087P}, OGLE14-073\cite{Terreran2017NatAs...1..713T}, and iPTF14hls\cite{Arcavi2017Natur.551..210A}. The error bars denote $1\sigma$ uncertainties. The light-curve phases of SN~2023vbw (Fig.~\ref{fig:LC}) are colour coded at the top. SN~2023vbw, OGLE14-073, and iPTF14hls all have higher velocities for longer durations compared to the mean SN~II and SN~1987A, with SN~2023vbw and iPTF14hls showing constant velocity evolution, similar to SN~1987A, during the optically thick (i.e., light-curve rise) phase.
\label{EDfig:vel}
}
\end{EDfigure}

\newpage

\begin{EDfigure}
\centering
\includegraphics[width=0.99\textwidth]{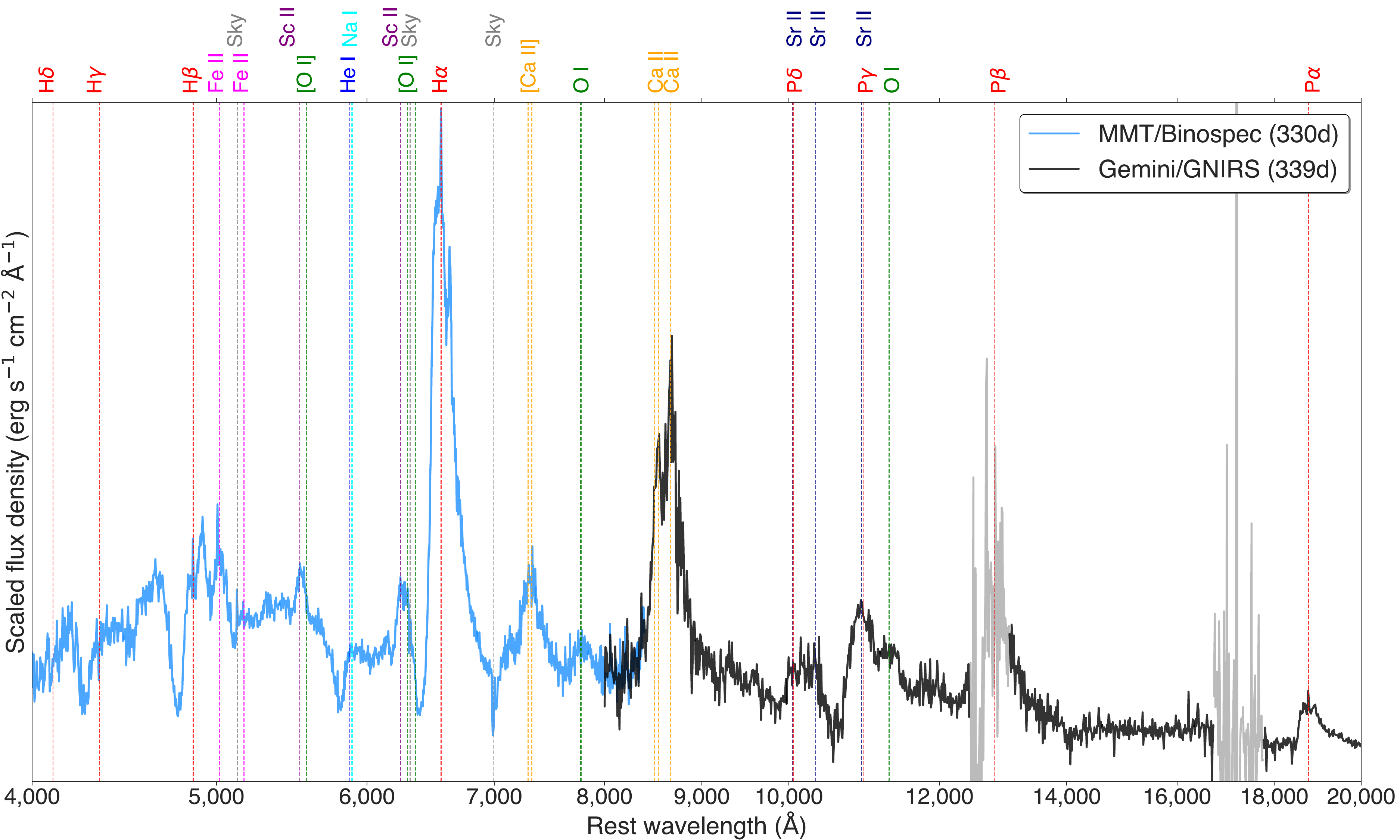}
\caption{
\textbf{Coeval optical and near-infrared spectra.}
The optical Binospec and NIR GNIRS spectra during the light-curve tail phase. The flux in the NIR telluric regions is masked with grey. The isolated Paschen lines (i.e., P$\alpha$ and possibly P$\gamma$) show double-peaked line profiles similar to those of H$\alpha$ and H$\beta$.
\label{EDfig:nir}
}
\end{EDfigure}

\newpage

\begin{EDfigure}
\centering
\includegraphics[width=0.99\textwidth]{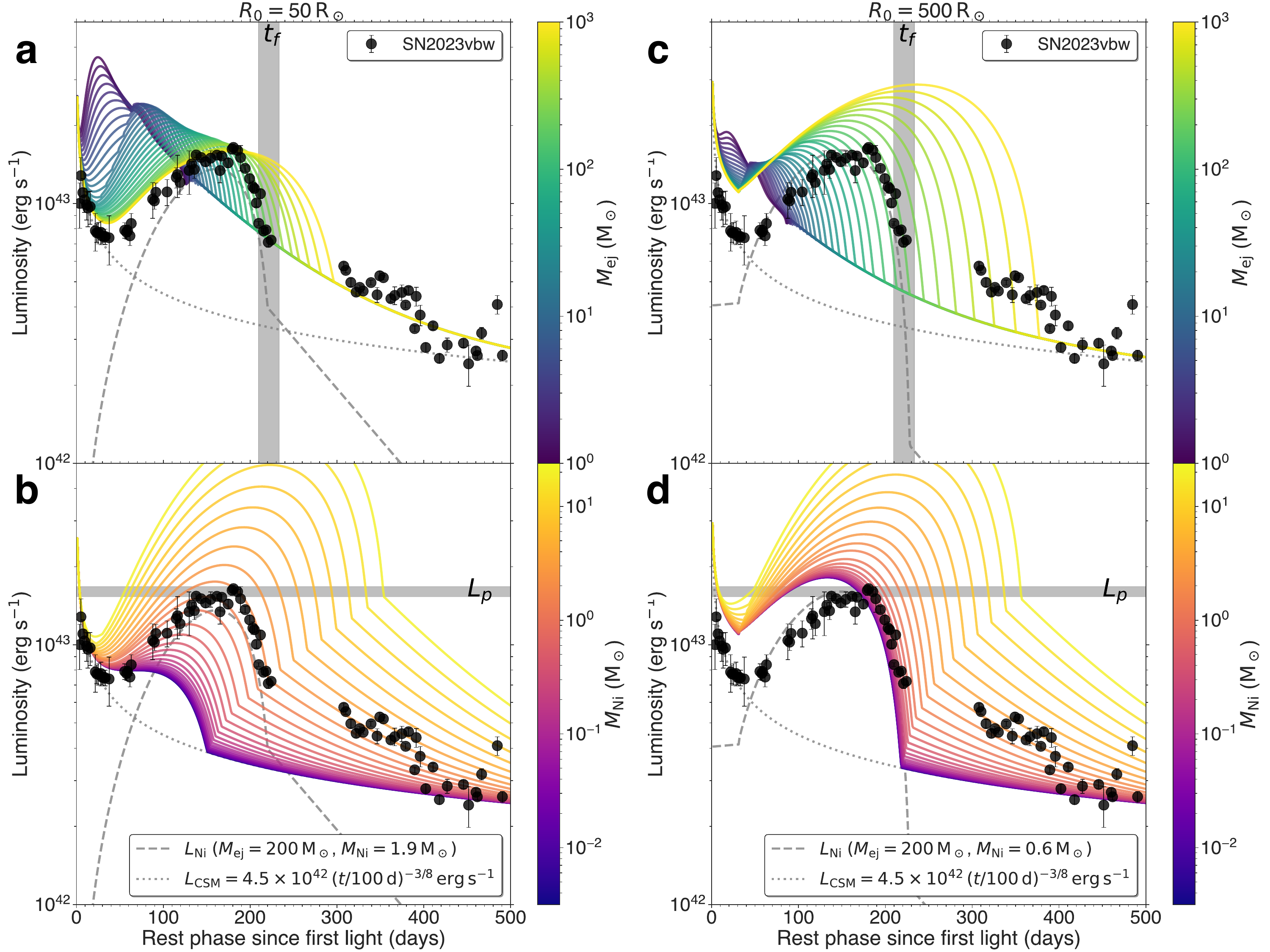}
\caption{
\textbf{Light-curve model dependence on the progenitor radius $R_0$.}
\textbf{a, b,} The $^{56}$Ni+CSM light-curve models colour-coded by ejecta masses ($M_{\rm ej}$; \textbf{a}) and $^{56}$Ni masses ($M_{\rm Ni}$; \textbf{b}) with $R_0=50\,{\rm R}_\odot$ (``Light-curve modeling'' section).
\textbf{c, d,} Similar to (\textbf{a, b}), but with $R_0=500\,{\rm R}_\odot$.
Compared to our fiducial model with $100~{\rm R}_\odot$ (Fig.~\ref{fig:mod}), reducing $R_0$ has little effect on the light curve, but increasing $R_0$ raises the early-phase luminosity, resulting in a plateau-like light-curve morphology, contrary to the slow rise observed in SN~2023vbw. 
\label{EDfig:mod_var}
}
\end{EDfigure}

\newpage

\begin{EDfigure}
\centering
\includegraphics[width=0.99\textwidth]{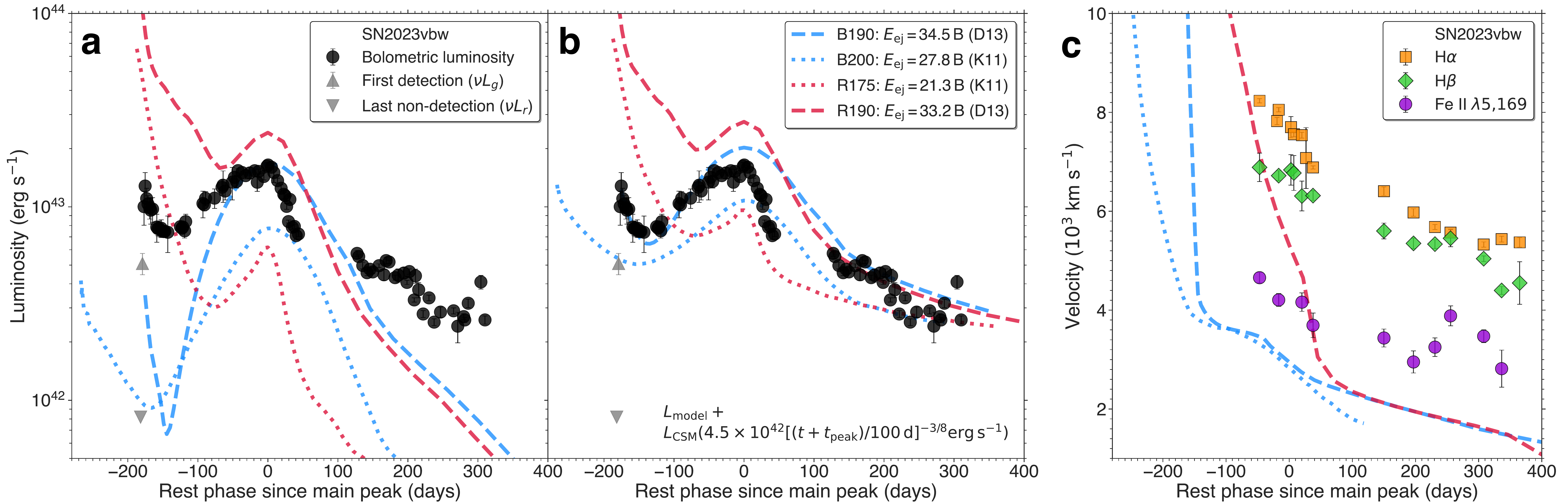}
\caption{
\textbf{Light-curve and velocity comparisons with numerical PISN models.}
\textbf{a, b,} The bolometric light curves of SN~2023vbw compared with the nearest neighbor PISN models (Fig.~\ref{fig:mod}) with available light curves (K11\cite{Kasen2011ApJ...734..102K} and D13\cite{Dessart2013MNRAS.428.3227D}): 
B190 (BSG with $M_{\rm ZAMS}=190\,{\rm M}_\odot$, $R_0=186\,{\rm R}_\odot$, $M_{\rm ej}=133.9\,{\rm M}_\odot$, $E_{\rm ej}=34.5\,{\rm B}$); 
B200 (BSG with $M_{\rm ZAMS}=200\,{\rm M}_\odot$, $R_0=94.4\,{\rm R}_\odot$, $M_{\rm ej}=200\,{\rm M}_\odot$, $E_{\rm ej}=27.8\,{\rm B}$);
R175 (RSG with $M_{\rm ZAMS}=175\,{\rm M}_\odot$, $R_0=2501\,{\rm R}_\odot$, $M_{\rm ej}=163.8\,{\rm M}_\odot$, $E_{\rm ej}=21.3\,{\rm B}$);
and R190 (RSG with $M_{\rm ZAMS}=190\,{\rm M}_\odot$, $R_0=4044\,{\rm R}_\odot$, $M_{\rm ej}=164.1\,{\rm M}_\odot$, $E_{\rm ej}=33.2\,{\rm B}$) where ${\rm B}=10^{51}\,{\rm erg}$, 
without (\textbf{a}) and with (\textbf{b}) the addition of the same CSM contribution estimated for SN~2023vbw (``Light-curve modeling'' section). The downward and upward gray triangles show the specific luminosities of the last nondetection and first detection, respectively (Fig.~\ref{fig:LC}). The phase is shifted with respect to the main peak for clarity. While the RSG models with large $R_0$ show luminous and extended shock-cooling phase that is inconsistent with the last nondetection, the BSG models with CSM interaction show the overall light-curve morphology that is consistent with SN~2023vbw, but with longer durations. 
\textbf{c,} Similar to (\textbf{a, b}), but for the line velocity evolution of SN~2023vbw (Extended Data Fig.~\ref{EDfig:vel}) compared with the photospheric velocity evolution of available models. The BSG models match the overall Fe~{\sc ii}\,$\lambda 5{,}169$ evolution of SN~2023vbw better than the RSG model, but with lower velocities, i.e., $(E_{\rm ej}/M_{\rm ej})^{1/2}$, which is responsible for the longer light-curve durations.
\label{EDfig:PISN_mod}
}
\end{EDfigure}
 

\end{addendum}





%













\end{document}